\documentclass[aps,prd,superscriptaddress,nofootinbib,twocolumn,notitlepage,longbibliography,10pt,floatfix]{revtex4-2}

\usepackage{amsmath,amssymb,amsbsy,amstext,amsfonts,mathrsfs,latexsym,bm,physics,graphicx,xcolor}
\usepackage[colorlinks=true,urlcolor=blue,anchorcolor=blue,citecolor=blue,linkcolor=blue,linktocpage=true,pdfa=true]{hyperref}
\usepackage{orcidlink}

\def\sqrtb{\mathpalette\DHLhksqrt}
\def\DHLhksqrt#1#2{%
\setbox0=\hbox{$#1\sqrt{#2\,}$}\dimen0=\ht0
\advance\dimen0-0.2\ht0
\setbox2=\hbox{\vrule height\ht0 depth -\dimen0}%
{\box0\lower0.4pt\box2}}

\newcommand{\GN}{G_{\scriptscriptstyle\mathrm{N}}}
\newcommand{\mpl}{M_{\scriptscriptstyle\mathrm{Pl}}}
\newcommand{\tb}{t_{\mathrm{b}}}
\newcommand{\xb}{x_{\mathrm{b}}}
\newcommand{\DE}{{\scriptscriptstyle\mathrm{DE}}}

\begin{document}

\title{How Much NEC Breaking Can the Universe Endure?}

\author{Elly Moghtaderi\,\orcidlink{0009-0008-4465-2007}}
\affiliation{Department of Applied Mathematics and Waterloo Centre for Astrophysics, University of Waterloo, Waterloo, ON N2L 3G1, Canada}
\affiliation{Perimeter Institute for Theoretical Physics, Waterloo, ON N2L 2Y5, Canada}

\author{Brayden R. Hull\,\orcidlink{0009-0004-4269-5354}}
\affiliation{Department of Applied Mathematics and Waterloo Centre for Astrophysics, University of Waterloo, Waterloo, ON N2L 3G1, Canada}
\affiliation{Perimeter Institute for Theoretical Physics, Waterloo, ON N2L 2Y5, Canada}

\author{Jerome Quintin\,\orcidlink{0000-0003-4532-7026}}
\affiliation{Department of Applied Mathematics and Waterloo Centre for Astrophysics, University of Waterloo, Waterloo, ON N2L 3G1, Canada}
\affiliation{Perimeter Institute for Theoretical Physics, Waterloo, ON N2L 2Y5, Canada}

\author{Ghazal Geshnizjani\,\orcidlink{0000-0002-2169-0579}}
\affiliation{Perimeter Institute for Theoretical Physics, Waterloo, ON N2L 2Y5, Canada}
\affiliation{Department of Applied Mathematics and Waterloo Centre for Astrophysics, University of Waterloo, Waterloo, ON N2L 3G1, Canada}

\begin{abstract}
Quantum fields can notoriously violate the null energy condition (NEC). In a cosmological context, NEC violation can lead to, e.g., dark energy at late times with an equation-of-state parameter smaller than $-1$ and nonsingular bounces at early times. However, it is expected that there should still be a limit in semiclasssical gravity to how much `negative energy' can accumulate over time and in space as a result of quantum effects. In the course of formulating quantum-motivated energy conditions, the smeared null energy condition has emerged as a recent proposal. This condition conjectures the existence of a semilocal bound on negative energy along null geodesics, which is expected to hold in semiclassical gravity. In this work, we show how the smeared null energy condition translates into theoretical constraints on NEC-violating cosmologies. Specifically, we derive the implied bounds on dark energy equation-of-state parameters and an inequality between the duration of a bouncing phase and the growth rate of the Hubble parameter at the bounce. In the case of dark energy, we identify the parameter space over which the smeared null energy condition is consistent with the recent constraints from the Dark Energy Spectroscopic Instrument (DESI).
\end{abstract}

\maketitle

\section{Introduction}

The role of energy conditions in classical general relativity (GR) is to determine what kind of matter fields one may `reasonably' include in the theory and, correspondingly, what kind of spacetime geometry one may physically expect. Historically, these conditions were formulated to represent the non-negativity of the energy densities or fluxes measured locally by observers, and they played three major roles in GR. To start with, under the assumption that some energy conditions hold, there exist rigorous singularity theorems (e.g., \cite{Penrose:1964wq,Hawking:1967ju}) that point to the breakdown of classical GR. Second, they have been used to forbid traversable wormhole solutions (e.g., \cite{Friedman:1993ty}) and third to prove positive asymptotic mass theorems for isolated objects (e.g., \cite{Schoen:1979zz}). Notably, the strong energy condition was initially regarded as one of the `reasonable' classical energy conditions. However, it is now generally accepted that it has been violated in the recent history of the cosmos by the current accelerated expansion of the universe (e.g., \cite{SupernovaSearchTeam:1998fmf,SupernovaCosmologyProject:1998vns,Peebles:2002gy,SupernovaCosmologyProject:2003dcn,SupernovaSearchTeam:2004lze,Frieman:2008sn}) and plausibly previously in the early universe, assuming an inflationary phase.

In the context of semiclassical gravity, determining what constitutes `reasonable' energy conditions becomes even more subtle, as the right-hand side of the Einstein equations now represents the quantum expectation value of the energy-momentum tensor. In fact, in regimes where singularities are predicted in classical GR, one expects quantum effects to change the game. For one, the hope is that a full theory of quantum gravity should be free of physically pathological singularities altogether. However, a more immediate impact of coupling quantum fields to gravity is that in quantum field theory (QFT) the energy densities at a point can be undoubtedly negative, and one can find solutions such that any of the local classical energy conditions are violated. A good example comes from the null energy condition (NEC), the weakest of these conditions, which plays a crucial role in GR theorems (see, e.g., \cite{Hawking:1973uf}): when considering quantum fields in flat or curved spacetime, the NEC can notoriously be violated (see, e.g., \cite{Casimir:1948dh,Epstein:1965zza,Brown:1969na,Wu:1986zz,Krommydas:2017ydo}). In fact, the violation of the NEC has also been shown to be very generic in the context of Hawking evaporation (e.g., \cite{Visser:1996iw,Visser:1996iv,Visser:1996ix,Visser:1997sd,Freivogel:2014dca,Bardeen:2017ypp,Kontou:2023ntd}), which is one of the most robust (theoretical) results in semiclassical gravity. Therefore, the relevant question is not whether the NEC holds at the fundamental level\footnote{In particular, NEC-violating quantum fields do not have to be `exotic' --- this can already happen within the standard model of particle physics and, for instance, could lead to nonpathological traversable wormholes \cite{Maldacena:2018gjk}.} locally, but rather if there is any bound on its violation. More specifically, physically speaking we do not expect the NEC to be violated systematically all the time and everywhere, arbitrarily. There should still exist some quantum or semiclassical energy conditions that tell us to what extent the NEC can be broken. 

The averaged null energy condition (ANEC; e.g., \cite{Borde:1987qr,Klinkhammer:1991ki,Ford:1994bj,Fewster:2006uf,Graham:2007va,Kontou:2015bta}) is one of the first proposals for a quantum-inspired, semiclassical version of the NEC. The ANEC stipulates that the quantum expectation value of the null energy, when averaged over an entire null (achronal) geodesic, is non-negative. It has been shown to hold for a wide class of QFTs in curved spacetimes and finds motivation from various approaches to quantum gravity (see, e.g., \cite{Wald:1991xn,Flanagan:1996gw,Verch:1999nt,Wall:2009wi,Kontou:2012ve,Kelly:2014mra,Kontou:2015yha,Faulkner:2016mzt,Hartman:2016lgu}). By definition, the ANEC is a global concept, though. Thus, it is difficult to apply the ANEC without proper knowledge of the entire spacetime, and correspondingly, it cannot always be used as a way of constraining local (or semilocal) violations of the NEC.

A more recent quantum-motivated semilocal null energy condition that has been conjectured to hold at the level of semiclassical gravity (or even further into a quantum gravity perturbative regime) is known as the smeared null energy condition (SNEC; \cite{Freivogel:2018gxj}). The SNEC is similar to the ANEC, except that the standard integral average is replaced by a weighted average through a smearing function. The goal of the smearing is to act in a similar fashion to a window function to regulate the energy condition in a particular regime, specifically a regime where the NEC would normally be violated. The SNEC then puts a bound on how much (semilocal) NEC breaking may be allowed within semiclassical gravity (i.e., within the regime of validity of the effective field theory). The SNEC can be heuristically understood as follows: quantum mechanically, we have Heisenberg's uncertainty principle, which can be stated as $\Delta E\,\Delta t\geq\hbar/2$, i.e., if we perform an observation with a time spread $\Delta t$, then we expect fluctuations in the energy of the measurement as $\Delta E\geq \hbar/(2\Delta t)$. Therefore, if one (a classical `observer') makes a measurement of the energy over a short smearing time, one may have large energy fluctuations that lead to negative energy, and the SNEC is a conjectured bound on how much negative energy one could actually observe given the coupling to gravity.\footnote{As we will see, the smearing is technically done along null geodesics (since this pertains to the \emph{null} energy condition). Still, we will also see in what sense this can be related in cosmology to a smearing with respect to time or the scale factor.} In other words, it is a form of `energy-spacetime uncertainty relation.'

The SNEC has passed a number of tests at the semiclassical level, has been generalized in various ways, and has led to novel singularity theorems with implications for black hole and cosmological spacetimes (see \cite{Freivogel:2020hiz,Fliss:2021gdz,Fliss:2021phs,Fliss:2023rzi,Fragoso:2024egu,Fliss:2024dxe}). For example, evaporating black holes satisfy the SNEC \cite{Freivogel:2020hiz,Kontou:2021lak}. Yet, there are many contexts in which the implications of the SNEC have not been explored, both in cosmological and black hole spacetimes. This work solely focuses on the former. Specifically, there are two main regimes in cosmology where the classical NEC might appear to be broken, but where a quantum modified version, like the SNEC, should hold: at late times and low energies; and at early times and high energies. The first regime is where we expect some form of dark energy (DE) to dominate the universe (e.g., \cite{SupernovaSearchTeam:1998fmf,SupernovaCosmologyProject:1998vns,Peebles:2002gy,SupernovaCosmologyProject:2003dcn,SupernovaSearchTeam:2004lze,Frieman:2008sn}). If this DE is of `phantom type' (meaning it has an equation-of-state [EoS] parameter smaller than $-1$; see, e.g., \cite{Caldwell:1999ew,Carroll:2003st,Caldwell:2003vq,Singh:2003vx,Dabrowski:2003jm,Elizalde:2004mq,Vikman:2004dc,Hu:2004kh,Nesseris:2006er}), then the classical NEC would be violated. The second regime pertains to the evolution of the universe prior to the start of standard big bang cosmology (and potentially before inflation if there was a phase of inflationary expansion). Indeed, in that regime, if a big bang singularity were avoided, then it might be that the universe underwent a period of NEC violation, resulting in, e.g., a big bounce (see, e.g., \cite{Novello:2008ra,Cai:2012va,Battefeld:2014uga,Brandenberger:2016vhg}).

Since phantom DE and a bouncing universe are both speculations --- though both lead to observable predictions that are testable --- we take the stance in this work that the SNEC might hold fundamentally and apply it to these NEC-violating spacetimes as a way of theoretically constraining them. Naturally, the SNEC remains a conjecture (at the same level as, e.g., the swampland conjectures in string theory), but it is nevertheless informative to explore its implications should the conjecture hold. In this work, we put aside theoretical issues that specific models may have related to having stable and consistent NEC violation in cosmology (such issues already put constraints on NEC-violating models; see, e.g., \cite{Cline:2003gs,Hsu:2004vr,Buniy:2005vh,Dubovsky:2005xd,Emparan:2005gg,Creminelli:2006xe,Buniy:2006xf,Arefeva:2006ido,Nicolis:2009qm,Garriga:2012pk,Sawicki:2012pz,Cline:2023hfw,Cline:2023cwm,Cline:2024zhs,Muralidharan:2024hsc,Easson:2013bda,Elder:2013gya,Rubakov:2014jja,Chatterjee:2015uya,Libanov:2016kfc,Kobayashi:2016xpl,Akama:2017jsa,Creminelli:2016zwa,Cai:2016thi,Cai:2017tku,deRham:2017aoj,Dobre:2017pnt,Banerjee:2018svi,Cai:2022ori,Volkova:2024mbn}). We simply assume that the NEC can be violated and see what constraints the SNEC would put on such cosmologies.

\paragraph*{Outline} We start in Sec.~\ref{sec:review} with a more thorough review of classical and semiclassical energy conditions, putting emphasis on the SNEC. Section \ref{sec:SNECcosmo} is devoted to the formulation of the SNEC in a cosmological background. This is applied to DE in Sec.~\ref{sec:DE}, covering the cases of a constant and time-dependent EoS. In particular, we compared our constraints with current observational bounds from the Dark Energy Spectroscopic Instrument (DESI; \cite{DESI:2024mwx,DESI:2024hhd,DESI:2025zgx,DESI:2025kuo}). Section \ref{sec:bounce} is devoted to exploring the constraints that the SNEC puts on nonsingular bouncing cosmologies. We end with a discussion in Sec.~\ref{sec:conclusions}.

\paragraph*{Conventions} Spacetime indices are denoted by Greek letters and are contracted with the spacetime metric tensor $g_{\mu\nu}$. We use the mostly positive metric signature and work with units where the speed of light $c$ and the reduced Planck constant $\hbar$ are set to unity.

\section{Review of some energy conditions}\label{sec:review}

\subsection{Classical versions}

General relativity is a classical theory of gravity. It relates the spacetime geometry (described by the metric tensor $g_{\mu\nu}$) to the energy content in the universe (described by the energy-momentum tensor $T_{\mu\nu}$). From the metric, one has knowledge of the curvature through the Riemann tensor $R^\alpha{}_{\mu\beta\nu}$, the Ricci tensor $R_{\mu\nu}\equiv R^{\alpha}{}_{\mu\alpha\nu}$, the Ricci scalar $R\equiv R^\mu{}_\mu$, and the Einstein tensor $G_{\mu\nu}\equiv R_{\mu\nu}-(R/2)g_{\mu\nu}$. The relation between $g_{\mu\nu}$ and $T_{\mu\nu}$ is known as the Einstein field equations,
\begin{equation}
    G_{\mu\nu}=8\pi \GN T_{\mu\nu}\,,\label{eq:EE}
\end{equation}
where $\GN$ is Newton's gravitational constant, or equivalently, taking the trace of \eqref{eq:EE} to replace $R$ by $T\equiv T^\alpha{}_\alpha$,
\begin{equation}
    R_{\mu\nu}=8\pi \GN \left(T_{\mu\nu}-\frac{1}{2}Tg_{\mu\nu}\right)\,.\label{eq:EET}
\end{equation}
In principle, one can get any desired spacetime geometry $g_{\mu\nu}$ by simply computing $G_{\mu\nu}$ from $g_{\mu\nu}$ and postulating an energy-momentum tensor $T_{\mu\nu}=G_{\mu\nu}/(8\pi \GN )$. Although this is mathematically legitimate, the resulting spacetime and its energy content may not be physically reasonable. On the other hand, if the equation of motion for a particular type of matter or the action leading to it is known, one can aim at solving it (given reasonable initial/boundary conditions) along with the Einstein equations to obtain both the spacetime geometry and the evolution of the matter degrees of freedom. However, ideally, we would like to obtain some generic features about possible physical spacetime geometries independent of a particular type of matter.

For these reasons, one usually adds energy conditions to the theory, i.e., reasonable conditions on the non-negativity of local energy densities or fluxes, formulated as linear combinations of components of $T_{\mu\nu}$, which in turn restrict the kind of geometries that are allowed. We call these `standard' energy conditions (null, strong, weak, dominant, etc.)~the \emph{classical energy conditions} (also known as pointwise energy conditions in the literature). For instance, the strong energy condition (SEC) states that, for any timelike vector $U^\mu$ (so $U^\mu U_\mu<0$),
\begin{equation}
    \left(T_{\mu\nu}-\frac{1}{2}Tg_{\mu\nu}\right)U^\mu U^\nu\geq 0\,,
\end{equation}
while the null energy condition (NEC) states that
\begin{equation}\label{eq:NECdef}
    T_{\mu\nu}K^\mu K^\nu\geq 0
\end{equation}
for any null vector $K^\mu$ (so $K^\mu K_\mu=0$).

One of the main motivations for the classical energy conditions comes from their utility in proving the singularity theorems (of Penrose and Hawking for instance). Such geometrical theorems always have some kind of `convergence condition' in their hypothesis, often conditions on $R_{\mu\nu}$ (akin to curvature bounds in mathematics); see, e.g., \cite{Senovilla:2014gza,Senovilla:2021pdg}. If one assumes GR, there is often a straightforward one-to-one relation between classical energy conditions and various convergence conditions (null, timelike, etc.); see, e.g., \cite{Hawking:1973uf,Poisson:2009pwt}.
For instance, one can prove the Penrose black hole singularity theorem \cite{Penrose:1964wq} assuming the null convergence condition (NCC), which says that $\forall\,K^\mu$ null,
\begin{equation}\label{eq:NCCdef}
    R_{\mu\nu}K^\mu K^\nu\geq 0\,.
\end{equation}
Similarly, the Hawking cosmological singularity theorem (e.g., \cite{Hawking:1967ju}) needs the timelike convergence condition (TCC), which says that $\forall\,U^\mu$ timelike, $R_{\mu\nu}U^\mu U^\nu\geq 0$. It is then straightforward to check that, whenever the Einstein field equations hold, NCC $\Leftrightarrow$ NEC, while TCC $\Leftrightarrow$ SEC. For the former equivalence, one uses \eqref{eq:EE}, while for the latter, one uses \eqref{eq:EET}.

If the gravitational theory is not GR, then things are a little more complicated. Nevertheless, modified theories of gravity are often (though not always) extensions of GR, e.g., $f(R)$ gravity, quadratic gravity, scalar-tensor gravity (such as $k$-essence, Galileon, Horndeski, etc.), and more. In that case (that is, when an Einstein frame exists), one can define an `effective' energy-momentum tensor that encapsulates the modifications to GR, such that the field equations read
\begin{equation} \label{eq:mod_field}
    G_{\mu\nu}=8\pi \GN T_{\mu\nu}^\mathrm{(eff)}\equiv 8\pi \GN \left(T_{\mu\nu}^\mathrm{(matter)}+T_{\mu\nu}^\mathrm{(ext)}\right)\,.
\end{equation}
Whether the extension of gravity is due to a scalar field, higher-curvature terms, or something else, one writes it `on the right-hand side' as $T_{\mu\nu}^\mathrm{(ext)}$. Then, the key question becomes whether the standard classical convergence conditions (NCC, TCC, ...), which are useful for singularity theorems, can be translated as before into `reasonable' energy conditions on the full $T_{\mu\nu}^\mathrm{(eff)}$, which includes $T_{\mu\nu}^\mathrm{(ext)}$. However, in these cases, determining what qualifies as `reasonable' energy conditions, even excluding quantum effects, is yet more ambiguous.

A straightforward way to see this is that if one has a modified theory that is written as \eqref{eq:mod_field}, i.e., if an appropriate Einstein frame exists (this is not always the case; see, e.g., \cite{Aoki:2018brq,DeFelice:2020eju,Aoki:2021zuy,DeFelice:2022uxv}), energy conditions may proceed in the `standard way' in that frame. However, the difficulty and challenges arise in the interpretation of the energy conditions across different frames (e.g., Einstein and Jordan frames) and if the energy conditions in the Einstein frame are physical. It is easy to have models where in one frame (Einstein to begin) there is satisfaction or violation of, e.g., the NEC, but where this result does not carry over into the Jordan frame identically (see \cite{Visser:1999de,Barcelo:2000zf} for an explicit example).

In general, to switch between the two frames one must apply the appropriate conformal (or even disformal) transformation on the metric and the additional fields present. In doing this, one obtains an `energy condition' in the Jordan frame, which possesses more terms than the Einstein-frame condition (see, e.g., \cite{Capozziello:2013vna,Capozziello:2014bqa}). For example, the NCC in the Einstein frame will read $R_{\mu \nu}^\mathrm{E}K^{\mu}K^{\nu}\geq 0$, where $R_{\mu \nu}^\mathrm{E}$ is the Einstein-frame Ricci tensor, i.e., the Ricci tensor computed from the Einstein-frame metric $g_{\mu\nu}^\mathrm{E}$. This condition does not imply $R_{\mu \nu}^\mathrm{J}K^{\mu}K^{\nu}\geq 0$, where the superscript refers to the Jordan frame this time and where the metrics across the two frames are related through a conformal transformation of the form $g_{\mu\nu}^\mathrm{E}=\Omega^2g_{\mu\nu}^\mathrm{J}$, for some spacetime-dependent function $\Omega^2$. There are also numerous subclasses of theories that have been investigated individually in regard to their energy conditions (e.g., \cite{Chatterjee:2012zh,Burger:2018hpz,Fliss:2023rzi}), though details can be theory dependent.

Despite the potential subtleties about energy conditions in modified gravity, some singularity theorems can nevertheless be rigorously proven in many modified gravity situations (e.g., \cite{Woolgar:2013yzk,Galloway:2013cea,Woolgar:2015wca,Alani:2016ail,Brown:2018hym,Kuipers:2019qby}) with certain adjustments to the convergence conditions to accommodate the modifications to gravity. Furthermore, there are also some singularity theorems in GR when energy conditions are weakened \cite{Senovilla:2006as,Fewster:2010gm,Fewster:2019bjg,Brown:2019orm,Fewster:2021mmz,Graf:2022mko} or when working in a nonsmooth setting \cite{Kunzinger:2014fka,Kunzinger:2015gwa,Graf:2017tli,Grant:2018jtw,Alexander:2019qcd,Graf:2019wuk,Kunzinger:2021das,Steinbauer:2022hvq,Cavalletti:2022phz,McCann:2023egg}.

\subsection{Semiclassical versions}

As already mentioned, classical GR breaks down in the approach to high-curvature regimes, in particular near classical singularities (as predicted under the aforementioned singularity theorems, which need to assume classical energy/convergence conditions). Therefore, we expect some form of quantum gravity to address that. However, even in low-curvature regimes far below Planck scales, the quantum effects of matter/tensor perturbations can be significant/observable (e.g., generating the seeds of large-scale structures, Hawking radiation, etc.), while the background spacetime is still well approximated by classical geometry. In the most conservative approach, we could apply semiclassical gravity to account for such effects by maintaining a `classical' spacetime geometry while quantizing the matter sector. In this case, assuming that the quantum fluctuations of the stress tensor are much smaller than its expectation value, the field equations would read\footnote{In cosmological perturbation theory, this picture becomes slightly more intricate since not only are the gravitational waves quantized, but the distinction between scalar metric perturbations and quantum modes corresponding to matter fields is not as clear-cut.}
\begin{equation}
    G_{\mu\nu}=8\pi \GN \langle\hat T_{\mu\nu}\rangle\,,\label{eq:semiGR}
\end{equation}
where $\langle\hat T_{\mu\nu}\rangle\equiv\langle\Psi|\hat T_{\mu\nu}|\Psi\rangle$ indicates the regulated quantum expectation value of the energy-momentum tensor quantum operator given some quantum state $|\Psi\rangle$. (See, e.g., \cite{Birrell:1982ix} for examples of how this may be computed.) One may then ask if there are reasonable energy conditions that $\langle\hat T_{\mu\nu}\rangle$ satisfies (in addition to those for remaining in the semiclassical approximation). Those would be semiclassical energy conditions. (There is vast literature on the subject, but see, e.g., \cite{Martin-Moruno:2017exc,Fewster:2012yh,Fewster:2017wmn,Kontou:2020bta} for reviews.) An example is the ANEC, which stipulates that
\begin{equation}
    \int_\gamma\dd\lambda\,\langle\Psi|\hat T_{\mu\nu}K^\mu K^\nu|\Psi\rangle\geq 0\label{eq:ANECdef}
\end{equation}
for any quantum state $|\Psi\rangle$ and for any achronal null geodesic $\gamma$. Note that $K^\mu$ in that case is the null tangent vector to $\gamma$, and $\lambda$ is an affine parameter along $\gamma$.
While the ANEC has been shown to hold for many physically motivated QFTs, its applicability is limited in the sense that verifying the ANEC requires knowledge of the full spacetime (and every null geodesic in that spacetime). The ANEC is a very nonlocal concept in that sense.

For this reason, we will be more interested in this work in semilocal energy conditions and specifically in a more recent proposal that is also applicable along null geodesics, namely the SNEC \cite{Freivogel:2018gxj,Freivogel:2020hiz}. The SNEC conjecture expands on the concept of quantum energy inequalities in $(1+1)$-dimensional (along timelike and null geodesics) and $(3+1)$-dimensional (along timelike geodesics) Minkowski spacetimes derived in \cite{Ford:1994bj} and further refined by \cite{Flanagan:1997gn}. It suggests introducing a real positive smearing function, $f_\tau^2(\lambda)$, where $\lambda$ is an affine parameter for a null geodesic $\gamma$ and $\tau$ denotes the smearing scale of the smearing function. The smearing function must be square integrable and normalized such that
\begin{equation}
    \int_\gamma\dd\lambda\,f_\tau^2(\lambda)=1\,.\label{eq:fnormalization}
\end{equation}
Then, the smearing of a test function $F(\lambda)$ along the null geodesic is done through
\begin{equation}
    \mathbb{E}_\tau[F]\equiv\int_\gamma\dd\lambda\,f_\tau^2(\lambda)F(\lambda)\,,
\end{equation}
and the corresponding characteristic smearing affine `length/time' scale $\tau$ is given according to
\begin{equation}
    \frac{1}{\tau^2}\equiv 4\int_\gamma\dd\lambda\,\left(\frac{\dd f_\tau}{\dd\lambda}\right)^2\,.\label{eq:taudef}
\end{equation}
As such, $f_\tau(\lambda)=\sqrtb{f_\tau^2(\lambda)}$ needs to be continuously differentiable, in principle.
The SNEC then claims that
\begin{equation}
    \mathbb{E}_\tau\left[\langle\Psi|\hat T_{\mu\nu}K^\mu K^\nu|\Psi\rangle\right]\geq -\frac{B}{\GN \tau^2}\label{eq:SNECdef}
\end{equation}
for any achronal null geodesic with null tangent vector $K^\mu$, any quantum state $|\Psi\rangle$ within the regime of perturbative quantum gravity, and any reasonable smearing function $f_\tau^2$ (more about this below). On the right-hand side, $B$ is a real positive constant, whose precise value is unknown (again, more about this later).

The original goal of the SNEC \cite{Freivogel:2014dca,Freivogel:2018gxj} was to find a general, (semi)local constraint on NEC violation in any dimensions that does not require additional quantities in QFT (such as entanglement entropy), with the motivation to restrict models (such as traversable wormholes or models with a large number of species that can lead to significant negative densities) that were not excluded by the ANEC or quantum inequalities in lower dimensions; see, e.g., \cite{Ford:1994bj,Flanagan:1997gn,Flanagan:2002bd,Fewster:2005gp}. As discussed in \cite{Freivogel:2018gxj}, there are similarities between the SNEC and other quantum inequalities (e.g., in $1+1$ dimensions \cite{Flanagan:1997gn,Flanagan:2002bd} and for classes of interacting conformal field theories \cite{Fewster:2004nj,Fewster:2018srj}). However, those results do not seem to extend to $3+1$ dimensions.

The initial SNEC formulation \cite{Freivogel:2018gxj} as described above, inspired by semilocal quantum inequalities \cite{Ford:1994bj}, uses smeared stress tensors to address large negative fluctuations of the energy-momentum tensor. To ensure this, and to mitigate the potential accumulated effect of Hawking radiation for null geodesics in the proximity of a black hole horizon, the regime of validity of the bound was assumed to be such that the smearing length scale $\tau$ is small compared to the background spacetime curvature scale. However, in the follow-up work \cite{Freivogel:2020hiz}, it was claimed that the bound may also hold for smearing length scales $\tau$ comparable to or larger than the spacetime curvature radius $\mathcal{R}\sim |R|^{-1/2}$ (or $\mathcal{R}\sim |R_{\mu\nu\alpha\beta}R^{\mu\nu\alpha\beta}|^{-1/4}$ if, e.g., the spacetime is Ricci flat), and this was shown for some explicit examples, such as evaporating black holes. As we will see in this work, the SNEC yields its most interesting constraint in cosmology when the smearing scale is of the order of the spacetime curvature.

An important property of the SNEC is that it reduces to the ANEC in the limit of infinite smearing scale. Indeed, taking $\tau\to\infty$, it is not too hard to see that \eqref{eq:SNECdef} reduces to \eqref{eq:ANECdef}. This was formally showed by \cite{Freivogel:2018gxj} with a rescaling method first used in \cite{Ford:1994bj,Fewster:2007ec,Brown:2018hym,Brown:2019orm}: as $\tau$ goes to infinity, the smearing function converges to a constant distribution and the SNEC reduces to the ANEC. Another nice aspect of SNEC is that it is invariant under rescaling of
the affine parameter. This addresses one of the arguments against the existence of null worldline inequalities in $3+1$ dimensions \cite{Fewster:2002ne}. Furthermore, \cite{Freivogel:2020hiz} imposes a Lorentz-invariant ultraviolet cutoff to show that the counterexample provided in \cite{Fewster:2002ne} also satisfies this bound.

The constant $B$ that appears on the right-hand side of \eqref{eq:SNECdef} is unknown. Its value would depend on the relation between the ultraviolet cutoff and Newton's constant. In the case of induced gravity on a brane in the anti-de Sitter/conformal field theory correspondence \cite{Leichenauer:2018tnq}, it was proven that $B=1/(32\pi)$. In what follows, we often use this value as a benchmark, but we must keep in mind that $B$ could potentially be as large as $1$, though more realistically it could also be much smaller. In fact, \cite{Freivogel:2020hiz} points to the fact that $B$ is typically expected to be much smaller than unity when semiclassical gravity is well under control.

Given the SNEC \eqref{eq:SNECdef} as a conjecture and assuming semiclassical gravity, namely \eqref{eq:semiGR}, one can say that classical geometries should satisfy a kind of `smeared null convergence condition',
\begin{equation}
    \mathbb{E}_\tau\left[R_{\mu\nu}K^\mu K^\nu\right]\geq-\frac{8\pi B}{\tau^2}\,.\label{eq:SNECgeodef}
\end{equation}
This is similar to the correspondence between the classical NEC and the NCC. In what follows, we wish to assume as little as possible about the nature of the quantum fields (or their specific quantum states) that would lead to violating the NEC. We simply assume they exist and drive the background cosmology so to yield, e.g., a phase of phantom DE acceleration or a bouncing cosmology. This allows us to do a purely `geometrical' analysis and ask: given a background cosmology with metric $g_{\mu\nu}$ and Ricci tensor $R_{\mu\nu}$, is the SNEC --- expressed as \eqref{eq:SNECgeodef} as a curvature bound --- satisfied or not? While geometrical, the bound \eqref{eq:SNECgeodef} nevertheless takes its root as a physical, semiclassical quantum inequality. Therefore, if a certain spacetime geometry cannot respect the SNEC \eqref{eq:SNECgeodef}, then we may say that the NEC-violating quantum fields (or states) that would presumably yield such a cosmological background are still nonphysical, as they violate a potentially fundamental energy condition, even though they may appear to be within the semiclassical approximation.

\section{The SNEC in cosmology}\label{sec:SNECcosmo}

\subsection{The full SNEC}

We want to explore the consequences of the SNEC in semiclassical GR, which following \eqref{eq:SNECgeodef} we express as
\begin{equation}
    \int_\gamma\dd\lambda\,f_\tau^2(\lambda)R_{\mu\nu}\frac{\dd X^\mu}{\dd\lambda}\frac{\dd X^\nu}{\dd\lambda}\geq-\frac{8\pi B}{\tau^2}\,,\label{eq:SNECgen1}
\end{equation}
where $\tau$ satisfies \eqref{eq:taudef}, and where $\gamma$ denotes a null geodesic parametrized by the curve $X^\mu(\lambda)$, so the tangent vector along the curve is $K^\mu\equiv\dd X^\mu/\dd\lambda$ such that $K^\mu K_\mu=0$. Note that the smearing function $f_\tau^2(\lambda)$ should have dimensions of mass, so that \eqref{eq:fnormalization} is respected. An example may be a Gaussian,
\begin{equation}
    f_\tau^2(\lambda)=\frac{1}{\sqrtb{2\pi}\,\tau}\exp(-\frac{(\lambda-\bar\lambda)^2}{2\tau^2})\,,\label{eq:Gaussian}
\end{equation}
or a Lorentzian,
\begin{equation}
    f_\tau^2(\lambda)=\frac{\tau}{\pi\sqrtb{2}\left((\lambda-\bar\lambda)^2+\tau^2/2\right)}\,,\label{eq:Lorentzian}
\end{equation}
for some real constants $\bar\lambda$ (controlling the center of the smearing) and $\tau>0$ (setting the smearing length scale). Both \eqref{eq:fnormalization} and \eqref{eq:taudef} hold in those cases over $\lambda\in(-\infty,\infty)$. Naturally for the Gaussian, the smearing scale $\tau$ also has the interpretation of the standard deviation of the distribution $f_\tau^2(\lambda)$ about its mean $\bar\lambda$.

Let us consider a flat Friedmann-Lema\^{i}tre-Robertson-Walker (FLRW) cosmology, described by the metric $g_{\mu\nu}\dd x^\mu \dd x^\nu=a(\eta)^2(-\dd\eta^2+\dd r^2+r^2\dd\Omega_{(2)}^2)$, where $a$ is the scale factor, $\eta$ is the conformal time related to the physical time through $\dd\eta\equiv a^{-1}\dd t$, and $\dd\Omega_{(2)}^2=\dd\theta^2+\sin^2(\theta)\dd\phi^2$ is the line element of the unit 2-sphere. An affine parameter $\lambda$ for a null geodesic $X^\mu(\lambda)$ can be taken such that $\dd\lambda\equiv a^2\dd\eta=a\,\dd t$ without loss of generality (see, e.g., \cite{Yoshida:2018ndv}). Then, we consider $K^\mu(\eta,r,\theta,\phi)=a^{-2}(1,1,0,0)$ as the generic (outgoing and nonrotating) null solution to the geodesic equation, $K^\nu\nabla_\nu K^\mu=\dd K^\mu/\dd\lambda+\Gamma^\mu_{\rho\sigma}K^\rho K^\sigma=0$. A simple calculation then shows
\begin{equation}
    R_{\mu\nu}K^\mu K^\nu=-2\frac{\dot H}{a^2}\,,\label{eq:RKK}
\end{equation}
where $H\equiv\dot a/a$ is the Hubble parameter and a dot denotes a derivative with respect to physical time, $t$. Thus, the SNEC \eqref{eq:SNECgen1} becomes
\begin{subequations}\label{eq:SNECFLRW0}
\begin{align}
    &\int_{\lambda_\mathrm{i}}^{\lambda_\mathrm{f}}\dd\lambda\,f_\tau^2(\lambda)\frac{\dot H}{a^2}\leq \frac{4\pi B}{\tau^2}\label{eq:SNECFLRW1}\\
    \Leftrightarrow & \int_{\lambda_\mathrm{i}}^{\lambda_\mathrm{f}}\dd\lambda\,\left(16\pi B\left(\frac{\dd f_\tau}{\dd\lambda}\right)^2-f_\tau^2(\lambda)\frac{\dot H}{a^2}\right)\geq 0\,,\label{eq:SNECFLRW2}
\end{align}
\end{subequations}
where one should view $\dot H/a^2$ as a function of $\lambda$. Note that we do not necessarily assume geodesic completeness here: if the null geodesic $\gamma$ is past (future) incomplete, then $\lambda_\mathrm{i}>-\infty$ ($\lambda_\mathrm{f}<\infty$); $\lambda_\mathrm{i}=-\infty$ ($\lambda_\mathrm{f}=\infty$) otherwise.

Given a solution for the scale factor, $a(t)$, one can express the affine length of null geodesics as
\begin{equation}
    \lambda(t)=\lambda_0+\int_{t_0}^t\dd\tilde t\,a(\tilde t)\,,\label{eq:lambdaoftasinta}
\end{equation}
where one chooses the arbitrary reference scale $\lambda(t_0)=\lambda_0$. From this, one can reexpress the integral in the SNEC as a time integral,
\begin{equation}
    \int_{t_\mathrm{i}}^{t_\mathrm{f}}\dd t\,f_\tau^2(\lambda(t))\frac{\dot H}{a}\leq\frac{4\pi B}{\tau^2}\,,\label{eq:SNECtime}
\end{equation}
where it is understood that $\lambda(t_\mathrm{i})=\lambda_\mathrm{i}$ and $\lambda(t_\mathrm{f})=\lambda_\mathrm{f}$. Alternatively, it is sometimes easier to find a solution for the Hubble parameter as a function of the scale factor, $H(a)$. In such a case, $\dot H=aH\dd H/\dd a$ and $\dd\lambda=\dd a/H$, so the SNEC becomes
\begin{equation}
    \int_{a_\mathrm{i}}^{a_\mathrm{f}}\frac{\dd a}{a}f_\tau^2(\lambda(a))\frac{\dd H}{\dd a}\leq\frac{4\pi B}{\tau^2}\,,\label{eq:SNECagen1}
\end{equation}
where
\begin{equation}
    \lambda(a)=\lambda_0+\int_{a_0}^a\frac{\dd\tilde a}{H(\tilde a)}\,,\label{eq:lambdaofagen}
\end{equation}
so $\lambda(a_\mathrm{i})=\lambda_\mathrm{i}$ and $\lambda(a_\mathrm{f})=\lambda_\mathrm{f}$. While \eqref{eq:SNECtime} and \eqref{eq:SNECagen1} show that we can perform changes of variable and express the SNEC integral as a smearing over time or scale factor, we should keep in mind that, physically, the smearing is really done along a null geodesic (as opposed to a timelike geodesic), as it is defined and normalized in terms of the null affine parameter $\lambda$.

Clearly, if the classical NEC --- or equivalently the NCC \eqref{eq:NCCdef} --- is satisfied in flat FLRW, then we know from \eqref{eq:RKK} that $\dot H(t)\leq 0$, hence the SNEC \eqref{eq:SNECFLRW1} is always satisfied. (This is always true, not just in FLRW, i.e., NEC $\Rightarrow$ SNEC, but SNEC $\nRightarrow$ NEC.) It thus becomes interesting to use the SNEC to constrain cosmologies that predict $\dot H>0$, i.e., violation of the classical NEC/NCC.

\subsection{The SNEC applied to individual matter components and the corresponding geometrical picture}

We have so far considered the SNEC for the total geometry of a FLRW universe, but the matter content of the universe consists of several fields, not all of which may violate the NEC in itself. Let us consider the case where the matter sector consists of several noninteracting components,
\begin{equation}
    T_{\mu\nu}=\sum_IT_{\mu\nu}^{I}\,,\label{eq:defTmunuI}
\end{equation}
where the index $I$ labels the different components (radiation, pressureless matter, DE, etc.). Then, we can define a `contribution to the curvature' for each matter component as
\begin{equation}
    G_{\mu\nu}^I\equiv 8\pi \GN T_{\mu\nu}^I\,,\label{eq:defGmunuI}
\end{equation}
such that the Einstein equations \eqref{eq:EE} with \eqref{eq:defTmunuI} and \eqref{eq:defGmunuI} give $G_{\mu\nu}=\sum_IG_{\mu\nu}^I$. Equivalently, we could define $R_{\mu\nu}^I\equiv 8\pi \GN (T_{\mu\nu}^I-(T^I/2)g_{\mu\nu})$, $R^I\equiv g^{\mu\nu}R_{\mu\nu}^I$, with $T^I\equiv g^{\alpha\beta}T_{\alpha\beta}^I$, such that $G_{\mu\nu}^I=R_{\mu\nu}^I-(R^I/2)g_{\mu\nu}$, $R_{\mu\nu}=\sum_IR_{\mu\nu}^I$, and $R=\sum_IR^I$. It is important to emphasize that we are not decomposing the Einstein equations into separate, independent equations here. Indeed, it would be inappropriate to separate the metric tensor into the sum of individual `metric components' because of the nonlinearities involved, hence $g_{\mu\nu}$ always appears as the `total' metric in the above.

The above definitions suggest we can view the source contributions to curvature as being the sum of several `components'. With this definition, the NEC for a given component, $T_{\mu\nu}^IK^\mu K^\nu\geq 0$, can now be expressed in terms of the corresponding `curvature contribution', $R_{\mu\nu}^IK^\mu K^\nu\geq 0$. If those hold for all components $I$, then naturally the full NEC and NCC, \eqref{eq:NECdef} and \eqref{eq:NCCdef}, are implied. However, the converse is not necessarily true, i.e., the validity of the full NEC does not mean that each component must individually respect the NEC. In other words, it is certainly possible to have some component violate the NEC, $T_{\mu\nu}^IK^\mu K^\nu\ngeq 0$, such that the full NEC \eqref{eq:NECdef} still holds.

Let us consider FLRW with a perfect fluid in $(t,x,y,z)$ Cartesian coordinates, so $g_{\mu\nu}=\mathrm{diag}(-1,a^2,a^2,a^2)$, $T^\mu{}_\nu=\mathrm{diag}(-\rho,p,p,p)$, and $T_{\mu\nu}=\mathrm{diag}(\rho,a^2p,a^2p,a^2p)$. With $K^\mu=(a^{-1},a^{-2},0,0)$ in these coordinates, we get
\begin{equation}
    T_{\mu\nu}K^\mu K^\nu=\frac{1}{a^2}(\rho+p)\,.
\end{equation}
This is in agreement with \eqref{eq:RKK} and the equation of motion one can derive from \eqref{eq:EE} in FLRW, namely $\dot H=-4\pi \GN (\rho+p)$. If the matter sector consists of several noninteracting components, then we may write $\rho=\sum_I\rho_I$ and $p=\sum_Ip_I$, hence $\dot H=-4\pi \GN \sum_I(\rho_I+p_I)$ and
\begin{equation}
    T_{\mu\nu}K^\mu K^\nu=\sum_IT_{\mu\nu}^IK^\mu K^\nu=\frac{1}{a^2}\sum_I(\rho_I+p_I)\,.
\end{equation}
Upon defining
\begin{equation}
    \dot H_I\equiv-4\pi \GN (\rho_I+p_I)\,,\label{eq:defHdotI}
\end{equation}
we get $\dot H=\sum_I\dot H_I$ and
\begin{equation}
    R_{\mu\nu}K^\mu K^\nu=\sum_IR_{\mu\nu}^IK^\mu K^\nu=-\frac{2}{a^2}\sum_I\dot H_I\,,
\end{equation}
so $\dot H$ separates into several $\dot H_I$ `curvature contributions'. As noted previously, the metric does not separate into `contributions', though, so in the above $a$ is really the `total' scale factor.

If we wish to impose the classical NEC on the $T_{\mu\nu}^I$s individually, then this is equivalent to imposing $\rho_I+p_I\geq 0$ for all $I$s. However, if we want to impose the SNEC on individual components, then things are a little different. Indeed, let us write the SNEC \eqref{eq:SNECdef} on individual components as
\begin{equation}
    \int_{\lambda_\mathrm{i}}^{\lambda_\mathrm{f}}\dd\lambda\,f_\tau^2(\lambda)T_{\mu\nu}^IK^\mu K^\nu\geq-\frac{B}{\GN\tau^2}\,,\label{eq:SNECTI}
\end{equation}
where we view $T_{\mu\nu}^I=\langle\hat T_{\mu\nu}^I\rangle$.
Then in FLRW this means
\begin{subequations}\label{eq:SNECcomponents01}
\begin{align}
    &\int_{\lambda_\mathrm{i}}^{\lambda_\mathrm{f}}\dd\lambda\,f_\tau^2(\lambda)\frac{\rho_I+p_I}{a^2}\geq-\frac{B}{\GN\tau^2}\label{eq:SNECcomponents0}\\
    \Leftrightarrow\quad&\int_{\lambda_\mathrm{i}}^{\lambda_\mathrm{f}}\dd\lambda\,f_\tau^2(\lambda)\frac{\dot H_I}{a^2}\leq\frac{4\pi B}{\tau^2}\,.\label{eq:SNECcomponents1}
\end{align}
\end{subequations}
Following from our assumption that there are no `energy transfers' (interactions) between components, we have $\nabla^\mu T_{\mu\nu}^I=0$ individually for any $I$s. This means $\dot\rho_I+3H(\rho_I+p_I)=0$, and for example if we consider their respective equations of state as $p_I=w_I\rho_I$ such that the $w_I$s remain constant with negligible fluctuations, then we get
\begin{equation}
    \rho_I=\rho_{I,0}a^{-3(1+w_I)}\,,\label{eq:rhoIsOfa} 
\end{equation}
hence from \eqref{eq:defHdotI},
\begin{equation}
    \dot H_I=-4\pi \GN (1+w_I)\rho_{I,0}a^{-3(1+w_I)}\,,\label{eq:dotHIcomp}
\end{equation}
where the constants $\rho_{I,0}$ indicate the value of the $\rho_I$s when $a=a_0=1$ (normalized to be the value of the scale factor in the universe today; henceforth the subscript $0$ indicates that a quantity is evaluated today). Recall $H$, the `total' Hubble parameter, is given by Friedmann's equation
\begin{equation}
    \frac{3}{8\pi \GN}H^2=\rho=\sum_I\rho_I=\sum_I\frac{\rho_{I,0}}{a^{3(1+w_I)}}\,.\label{eq:Friedmannconstraint}
\end{equation}
Furthermore, since $\dd\lambda=\dd a/H$, the SNEC \eqref{eq:SNECcomponents1} becomes
\begin{equation}
    L_I\equiv\int_{a_\mathrm{i}}^{a_\mathrm{f}}\frac{\dd a}{H}\,f_\tau^2(\lambda(a))\frac{\dot H_I}{a^2}\leq\frac{4\pi B}{\tau^2}\,.\label{eq:SNECcosmomulti}
\end{equation}
Therefore, the integral on the left-hand side of the inequality, denoted as $L_I$, may be expressed as
\begin{align}
    L_I=&-\sqrtb{6\pi \GN }(1+w_I)\rho_{I,0}\nonumber\\
    &\times\int_{a_\mathrm{i}}^{a_\mathrm{f}}\dd a\,\frac{f_\tau^2(\lambda(a))}{a^{5+3w_I}}\bigg(\sum_J\frac{\rho_{J,0}}{a^{3(1+w_J)}}\bigg)^{-1/2}\,,\label{eq:Lgen1}
\end{align}
when combining \eqref{eq:dotHIcomp} and \eqref{eq:Friedmannconstraint}, assuming an expanding universe with $H>0$.

Note that, while convenient, the whole `curvature contribution' terminology was not necessary here, i.e., one could simply express the SNEC in terms of the energy components, \eqref{eq:SNECcomponents0}, and use \eqref{eq:rhoIsOfa} and \eqref{eq:Friedmannconstraint} to arrive at \eqref{eq:Lgen1}. Nevertheless, what it shows is that when applying the SNEC to a single component $I$, one still has to take into account all other components [the sum over $J$ in \eqref{eq:Lgen1}]. This is because evaluating the SNEC requires knowledge of the whole cosmology. The above shall be very useful in cosmological contexts where several fields `compete' in the energy budget, as in the late era (see the next section). This will be less important for a bouncing universe since then a single field usually dominates the dynamics.

\section{Dark energy}\label{sec:DE}

The onset of the DE era, which marks the current acceleration of cosmic expansion, is a relatively recent event in the history of the universe. The primary candidate for explaining this acceleration is a cosmological constant $\Lambda$, characterized by $w=-1$, with the time variation of the DE EoS considered negligible on cosmological scales. Interestingly, recent observations (DESI; \cite{DESI:2024mwx,DESI:2024hhd,DESI:2024aqx,DESI:2024kob,DESI:2025zgx,DESI:2025kuo}) seem to support the assumption that the DE EoS may be time dependent. Consequently, current fitting models often use a constant $w$ and, to next order, a parametrized time-dependent EoS, assuming that its variation is small, which we will discuss in subsequent sections. Theoretically, beyond the cosmological constant, these phenomenological models may serve as reasonable approximations during brief cosmological phases, but they are less applicable for extended periods of evolution. Additionally, this paper focuses on exploring the implications of the SNEC on NEC-violating (phantom) DE driven by quantum effects, which are challenging to model and might reveal a more intricate time dependence in the EoS. Nevertheless, as a first step, our goal is not to construct such top-down models but rather to adopt an intermediate approach, where we assume these phenomenological models act as proxies for NEC violation (meaning the EoS variation is assumed to be minor, monotonic, and to occur over long durations) and explore the implications of the SNEC for such cosmologies. However, it would be intriguing to further investigate time-dependent equations of state and model NEC violation through top-down QFT constructions leading to phantom-like models.

\subsection{Constant equation of state}

For simplicity at first, and bearing in mind the caveats stated above, we consider a universe that has a phantom DE fluid with constant EoS parameter $w\equiv p_\DE/\rho_\DE<-1$ with the addition of pressureless matter (cold dark matter and baryons). As is evident from the sign of $-3(1+w_I)$ in \eqref{eq:rhoIsOfa} and as expected in an expanding universe, pressureless matter dominates at early times, whereas the phantom DE becomes dominant at later times. Denoting their respective fractional energy densities today by $\Omega_\DE\equiv 8\pi \GN \rho_{\DE,0}/(3H_0^2)$ and $\Omega_\mathrm{m}\equiv 8\pi \GN \rho_{\mathrm{m},0}/(3H_0^2)$, we have $\Omega_\mathrm{m}=1-\Omega_\DE$. Then, the SNEC integral \eqref{eq:Lgen1} applied to the DE component becomes
\begin{align}
    L_\DE=&-\frac{3}{2}H_0(1+w)\Omega_{\DE}^{1/2}\nonumber\\
    &\times\int_{a_\mathrm{i}}^{a_\mathrm{f}}\dd a\,\frac{f_\tau^2(\lambda(a))}{a^{(7+3w)/2}}\left(1+\frac{1-\Omega_\DE}{\Omega_{\DE}}a^{3w}\right)^{-1/2}\,,\label{eq:LDEw}
\end{align}
and the affine length \eqref{eq:lambdaofagen} satisfies
\begin{align}
    \lambda_0&-\lambda(a)=\frac{1}{H_0\Omega_\DE^{1/2}}\int_a^{a_0}\dd\tilde a\,\frac{a^{3(1+w)/2}}{\sqrtb{1+\frac{1-\Omega_\DE}{\Omega_\DE}a^{3w}}}\nonumber\\
    &=\frac{2\tilde a^{(5+3w)/2}}{(5+3w)H_0\Omega_\DE^{1/2}}\nonumber\\
    &\quad\times{}_2F_1\Big(\frac{1}{2},\frac{5+3w}{6w},\frac{5+9w}{6w},-\frac{1-\Omega_\DE}{\Omega_\DE}\tilde a^{3w}\Big)\bigg|_{\tilde a=a}^{a_0}\,,\label{eq:lambdaDE}
\end{align}
where ${}_2F_1$ is the Gaussian hypergeometric function. One can check that in the limit $a\to\infty$, $\lambda(a)$ reaches a finite constant if $w<-5/3$, indicating future null geodesic incompleteness (i.e., a big rip; e.g., \cite{Caldwell:2003vq,Nojiri:2005sx,Briscese:2006xu}), while null geodesics are complete if $w\geq -5/3$. As for the past, since pressureless matter becomes dominant one naturally reaches a big bang singularity taking the scale factor to zero, but the model can only be applied until matter-radiation equality when $a_\mathrm{eq}\approx 1/3400$. (Improving the model with the addition of a radiation component would not affect the subsequent results significantly, since what should matter most is the DE era at late times and where the NEC for DE is most motivated to be violated. In the same vein, what happens before standard big bang cosmology is irrelevant for this part.)

Let us evaluate \eqref{eq:LDEw}, first making some rough approximations to gain some intuition. Specifically, let us take a continuous uniform distribution for our smearing function as
\begin{equation}
    f^2_\tau(\lambda)=\begin{cases}
        1/(\lambda_2-\lambda_1) & \lambda\in[\lambda_1,\lambda_2]\\
        0 & \lambda\notin[\lambda_1,\lambda_2]\,.
    \end{cases}\label{eq:CUD}
\end{equation}
Note that such a smearing function is not quite appropriate since one cannot use \eqref{eq:taudef} to define the smearing scale $\tau$ (i.e., $f_\tau$ is not continuously differentiable).\footnote{One could instead work with a smooth version of the above (e.g., \cite{Weiss:2025}), but for the purpose of the approximation, \eqref{eq:CUD} is fine, and we will resort to other smooth smearing functions later.} However, it does respect \eqref{eq:fnormalization} as long as $\gamma$ covers any range $[\lambda_\mathrm{i},\lambda_\mathrm{f}]$ with $\lambda_\mathrm{i}\leq\lambda_1$ and $\lambda_2\leq\lambda_\mathrm{f}$. Then, since the above smearing function is symmetric about its mean, $\mathbb{E}_\tau[\lambda]=(\lambda_1+\lambda_2)/2$, we can take $\tau$ such that it corresponds to the standard deviation of the `probability' distribution itself, $\tau=\sqrtb{\mathbb{E}_\tau[\lambda^2]-\mathbb{E}_\tau[\lambda]^2}=\Delta\lambda/\left(2\sqrtb{3}\right)$, where $\Delta\lambda\equiv\lambda_2-\lambda_1$. Correspondingly, the SNEC \eqref{eq:SNECcosmomulti} using \eqref{eq:LDEw} becomes
\begin{align}
    \int_{a_1}^{a_2}\frac{\dd a}{a^{(7+3w)/2}\sqrtb{1+\frac{1-\Omega_\DE}{\Omega_\DE}a^{3w}}}\leq\frac{1}{H_0|1+w|\Omega_{\DE}^{1/2}\Delta\lambda}\,,\label{eq:SNECDEapprox}
\end{align}
where $\lambda(a_1)=\lambda_1$, $\lambda(a_2)=\lambda_2$, and for concreteness we take $B=1/(32\pi)$. The integral on the left-hand side is another hypergeometric function, so once more for the purpose of a rough approximation, let us perform an expansion about $\Omega_\DE\approx 1$, which is to say that DE should dominate over pressureless matter in the regime of interest. Then, to leading order one finds
\begin{equation}
    \frac{16|1+w|}{(5+3w)^2}\sinh^2\bigg(\frac{5+3w}{4}\ln\Big(\frac{a_2}{a_1}\Big)\bigg)\leq 1\,.
\end{equation}
What this equation and \eqref{eq:SNECDEapprox} show is that the bound on the EoS $w$ depends mainly on the chosen smearing scale $\tau\propto\Delta\lambda$, which controls $a_2/a_1$ in the above. For instance, if we set the final integration time to today, $a_2=a_0=1$, and the lower integration limit to be matter-radiation equality, $a_1=a_\mathrm{eq}$, then the SNEC is satisfied as long as $w\gtrsim-1.003$, while the bound is much weaker if the integration only begins when DE starts dominating, e.g., $w\gtrsim -3.58$ when $a_1=4/7$ (corresponding to a redshift of $0.75$). This is intuitively expected when assuming $w<-1$ persisted longer, because then NEC violation accumulates more and the bound should get tighter. Similarly, at fixed $a_1$, we always find that the SNEC can only hold if $w\geq -1$ if we send $a_2\to\infty$. Indeed, in such a limit ($\tau\to\infty$), we should recover the ANEC, which forbids the dominance of a NEC-violating phantom field forever in the future.

The above highlights an important fact about our implementation of the SNEC, which is that what matters is not only the effect of DE in the late-time universe when it initiates a period of accelerated expansion, but rather the presence of DE as a phantom field in the whole history of the observable universe. In our simple model with only matter [$\rho_\mathrm{m}(a)=\rho_{\mathrm{m},0}a^{-3}$] and DE [$\rho_\DE(a)=\rho_{\DE,0}a^{-3(1+w)}$] such that $\Omega_\mathrm{m}=1-\Omega_\DE$, the transition between matter domination and DE domination in terms of the DE EoS occurs at the scale factor value [the solution to $\rho_\mathrm{m}(a_\star)=\rho_\DE(a_\star)$]
\begin{equation}
    a_\star=\left(\frac{\Omega_\DE}{1-\Omega_\DE}\right)^{1/(3w)}\,.
\end{equation}
The corresponding redshift is $z_\star=1/a_\star-1$. As long as $\Omega_\DE>1/2$, then $a_\star<1$ for any $w<-1$, meaning that DE starts dominating at some point in the past. Note that the more negative $w$ is, the closer to today DE starts dominating, so the shorter the DE-dominated phase. If we evaluate \eqref{eq:SNECDEapprox} from the start of DE domination ($a_1=a_\star$) to today ($a_2=1$), then we find that the SNEC is always satisfied, for any EoS, no matter how negative (assuming the benchmark value of $B=1/(32\pi)$; we discuss the value of $B$ further below). The SNEC in this case could imply a constraint on $w$ if we integrated to the future ($a_2>1$; in fact, integrating all the way to $a_2\to\infty$ we would recover $w\nless -1$). However, we do not find it theoretically or observationally motivated since we cannot `observe' the future. In particular, it could be that DE decays to the future (see, e.g., \cite{Agrawal:2018own,Andrei:2022rhi,Lehners:2024qaw}). Therefore, it is more reasonable to consider $a_1<a_\star$. Prior to DE domination, our ability to observationally probe the DE EoS is much lower, as its effect on cosmology is much weaker. Nevertheless, if we assume that DE, as a constant EoS phantom field, was present for essentially as long as we observe the universe, then the SNEC provides interesting constraints, qualitatively in agreement with our analytical estimate obtained in \eqref{eq:SNECDEapprox}, to which we turn to next.

\begin{figure}[t]
    \centering
    \includegraphics[width=0.99\linewidth]{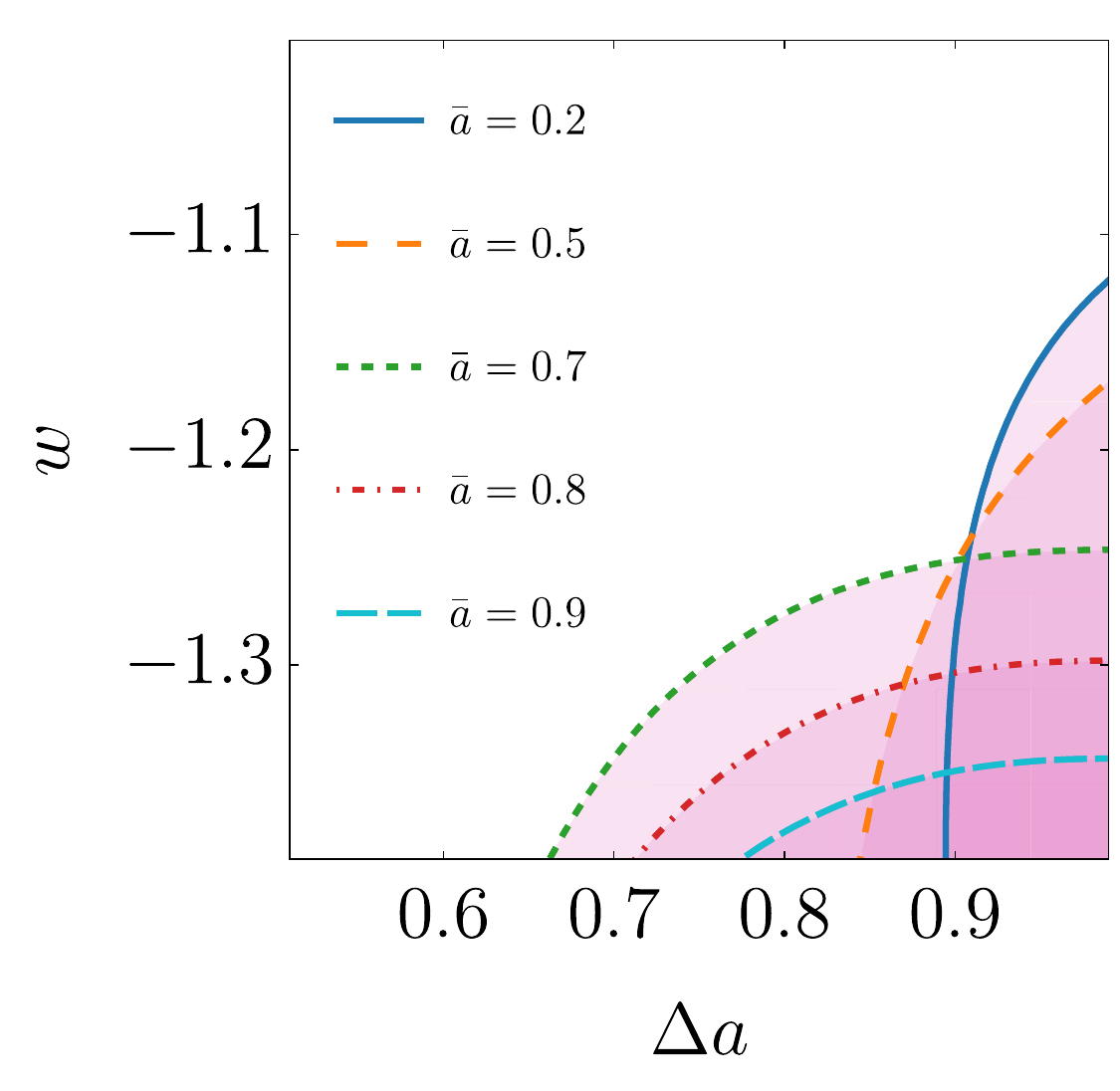}
    \caption{SNEC constraints on constant-EoS phantom DE. The constraint on the EoS $w$ is shown as a function of $\Delta a$ (related to the smearing width) for different values of $\bar a$ (related to the center of the smearing). The other parameters are fixed to $\Omega_\DE=0.7$ and $B=1/(96\pi)$. The pink regions are where the SNEC is violated.}
    \label{fig:abar}
\end{figure}

Approaching the problem numerically this time, let us take a smooth smearing function, specifically the Gaussian \eqref{eq:Gaussian}, though we found the results to be quantitatively similar for different choices, such as the Lorentzian smearing function \eqref{eq:Lorentzian}. We shall fix the lower and upper integration limits: we take $a_\mathrm{f}=a_0=1$ so that there is no contribution to the SNEC coming from the future as explained above; and the lower integration limit is taken as small as reasonable, so $a_\mathrm{i}=a_\mathrm{eq}$ for concreteness. We then properly normalize the Gaussian \eqref{eq:Gaussian} such that its integral from $\lambda_\mathrm{i}=\lambda(a_\mathrm{i})$ to $\lambda_\mathrm{f}=\lambda(a_\mathrm{f})$ is unity. The center of the Gaussian is parametrized according to $\bar\lambda=\lambda(\bar a)$, and likewise, the smearing scale is parametrized according to $\tau=\big(\lambda(a_+)-\lambda(a_-)\big)/2$, with $a_\pm=\bar a\pm\Delta a/2$ as long as $\bar a+\Delta a/2\leq a_0$ and $\bar a-\Delta a/2\geq a_\mathrm{eq}$. If $\bar a+\Delta a/2>a_0$, we instead set $a_+=a_0$ and $a_-=a_0-\Delta a$. Likewise if $\bar a-\Delta a/2<a_\mathrm{eq}$, we instead set $a_-=a_\mathrm{eq}$ and $a_+=a_\mathrm{eq}+\Delta a$. In that sense, $\bar a$ controls the position of the smearing in time, and $\Delta a$ (which is associated with the smearing scale) controls the total width. Recall that $\lambda(a)$ is given by \eqref{eq:lambdaDE}, where the value of $\lambda_0$ is arbitrary. Likewise, the numerical value of $H_0$ cancels out, so it is irrelevant.

\begin{figure*}[t]
    \centering
    \includegraphics[width=0.45\textwidth]{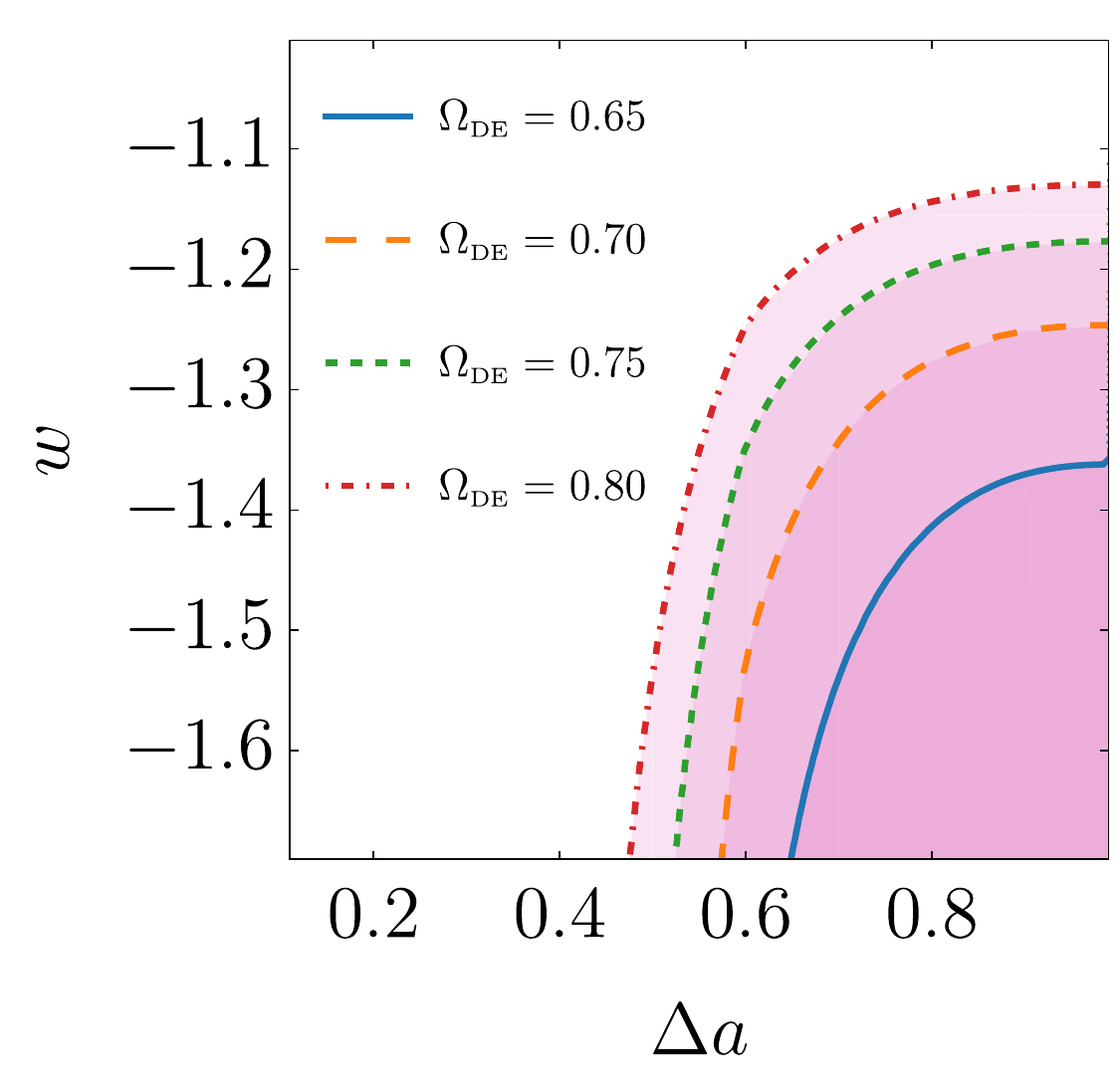}
    \hspace*{0.05\textwidth}
    \includegraphics[width=0.45\textwidth]{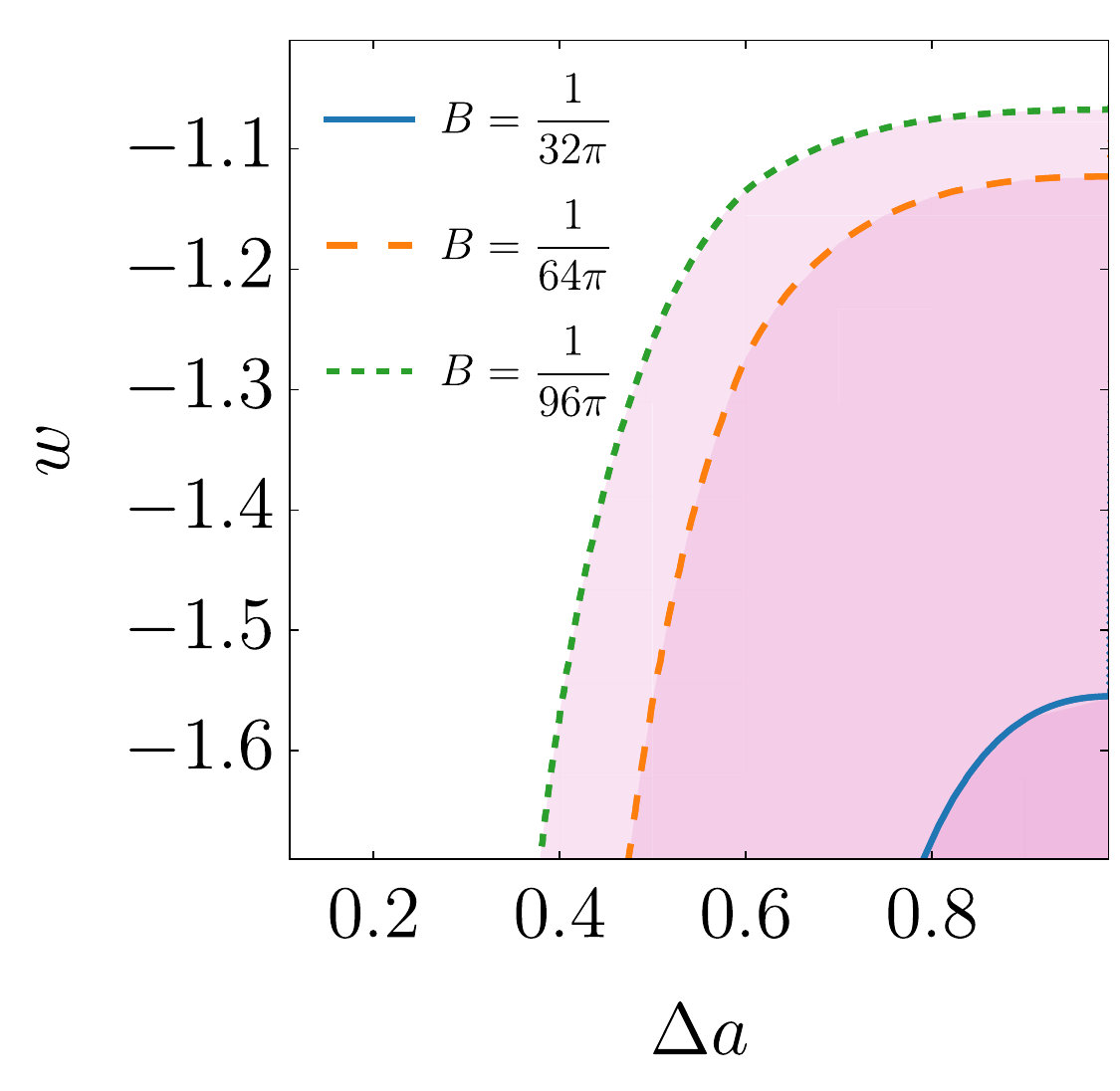}
    \caption{Same plots as Fig.~\ref{fig:abar}, except now in the left plot we fix $\bar a=0.7$, $B=1/(96\pi)$, and show different values of $\Omega_\DE$; and in the right plot we fix $\bar a=0.7$, $\Omega_\DE=0.9$, and show different values of $B$.}
    \label{fig:OmegaDE-B}
\end{figure*}

We can then numerically perform the integral of \eqref{eq:LDEw} and determine when the SNEC \eqref{eq:SNECcosmomulti} is satisfied or not. This depends on the value of $w$, $\Omega_\DE$, $\bar a$, $\Delta a$, and $B$. For certain choices of parameters, we determine the critical value of the DE EoS $w$, beneath which the SNEC is violated. A first example is shown in Fig.~\ref{fig:abar}, where the SNEC constraint on $w$ is plotted as a function of $\Delta a$, with different curves showing different values of $\bar a$. The other parameters are fixed as $\Omega_\DE=0.7$ and $B=1/(96\pi)$. Similar plots are shown in Fig.~\ref{fig:OmegaDE-B}, where $\Omega_\DE$ and $B$ are varied.

A first thing we observe is that the constraint on $w$ is always monotonically increasing in $\Delta a$. This was to be expected: the smearing scale enters both the `left- and right-hand sides' of the SNEC [expressed as, e.g., \eqref{eq:SNECTI}]. On the `left-hand side', the smearing scale controls the normalization of the smearing and how long NEC violation can accumulate around the peak of the smearing, but its main effect is on the `right-hand side', which scales as $1/\tau^2$. If $\Delta a$ is small, the smearing scale $\tau$ is correspondingly small, and as such, the SNEC permits large amounts of negative energy, hence $w$ can be very negative. Conversely, if $\Delta a$ is large, the smearing scale $\tau$ is correspondingly large, and as such, the SNEC allows only small amounts of negative energy; hence the constraint on $w$ is much closer to the NEC divide ($w=-1$). Consequently, the largest constraint (at fixed $\bar a$, $B$, and $\Omega_\DE$) always comes from the longest smearing scale allowed, which in our case corresponds to having $\Delta a=a_0-a_\mathrm{eq}\approx 0.999706$. This essentially amounts to the whole `observable' history of the universe, hence one can say that the smearing length scale is of the order of the `curvature radius' of the universe ($\mathcal{R}\sim|R|^{-1/2}\sim|H_0|^{-1}$). We recall that the original SNEC formulation \cite{Freivogel:2018gxj} assumed that the smearing length scale $\tau$ had to be small compared to the background spacetime curvature scale $\mathcal{R}$, but subsequent work \cite{Freivogel:2020hiz} found the SNEC could hold when $\tau$ is of the order of $\mathcal{R}$ (or even larger, as in the ANEC limit $\tau\to\infty$). In cosmology, assuming $B$ is not too small, the SNEC really appears to be most relevant as a constraint when $\tau$ is of the order of $\mathcal{R}$.

Under the premise that we want to smear only over `past observable scales', the maximal value of $\Delta a$ is naturally bounded by $a_0$ since we cannot observe the future. There is also a bound to the past, here conservatively taken to be $a_\mathrm{eq}$, though in principle we are able to observationally infer the cosmology with confidence all the way to much smaller $a$ values (e.g., down to big bang nucleosynthesis). At the same time, our inference of the DE EoS from observations is mostly dominated by late-time cosmology (see, e.g., \cite{Escamilla:2024fzq,Yang:2024kdo,DESI:2024aqx,DESI:2024kob,Sabogal:2024qxs,Liu:2024gfy,Ye:2024ywg,Yang:2025kgc,Keeley:2025stf,Ormondroyd:2025exu,Berti:2025,DESI:2025zgx,DESI:2025kuo,Ormondroyd:2025iaf} for recent observational constraints on DE as a function of redshift). In that sense, we cannot say that we observationally `measure' $w$ in a robust manner over a large smearing scale $\Delta a\sim 1$. Yet in principle when a constant EoS is assumed for DE for a long cosmological period, it still implies that DE is present at all times (even though its contribution to the background evolution is potentially negligible at early times), and it has a significant effect on determining the SNEC bound.

The other parameter that controls the smearing is $\bar a$; it dictates where $f_\tau^2(\lambda)$ is centered and thus where it peaks. Different values of $\bar a$ are presented in Fig.~\ref{fig:abar}, showing how it affects that SNEC constraint. The `amount of negative energy' entering the SNEC has a nontrivial $a$ dependence [cf.~\eqref{eq:LDEw}]; in fact, for some range $-2 \lesssim w < -1$ it tends to pick up significant contributions at smaller $a$ values (earlier times), hence the SNEC is more constraining the lower the center of the smearing function as it picks up more of that negative energy. At the same time, it might not be too realistic to set $\bar a$ to a very small value considering we observe DE best in the late universe ($z\lesssim 1$, so $a\gtrsim 0.5$). This is the motivation for fixing $\bar a=0.7$ in Fig.~\ref{fig:OmegaDE-B}, though we have to keep in mind that this is somewhat of an arbitrary choice simply to isolate the dependence on the other parameters ($\bar a\sim 0.7$ finds motivation in the case of a time-dependent EoS, which we discuss in the next subsection). Moreover, the SNEC --- as a conjecture --- states that within its regime of validity it should hold for all possible smearing functions, so in particular for any values of $\bar a$ and $\Delta a$. Under this interpretation, the most stringent constraint should be the one to hold. According to Fig.~\ref{fig:abar}, this would imply $w\gtrsim -1.12$ [for $\Omega_\DE=0.7$ and $B=1/(96\pi)$].

In the left plot of Fig.~\ref{fig:OmegaDE-B}, we see how the constraint varies depending on the relative energy density of DE today, $\Omega_\DE$. The trend is intuitively understood: the larger $\Omega_\DE$ is, the more DE there is in the universe over time, the stronger the constraint on $w$ upon applying the SNEC. The dependence on $B$, depicted in the right plot of Fig.~\ref{fig:OmegaDE-B}, is also straightforward to understand from \eqref{eq:SNECTI}: more `negative energy' is allowed the larger $B$ is, and this is what the right plot of Fig.~\ref{fig:OmegaDE-B} shows. Note that we choose an artificially large amount of DE in that plot, $\Omega_\DE=0.9$, in order to depict a sizable constraint from the case $B=1/(32\pi)$. Recall that this value of $B$ comes from some examples of quantum inequalities similar to the SNEC, which have been derived in different contexts (lower dimensions, conformal field theories, etc.). In the present situation, the constraints on phantom DE with $B=1/(32\pi)$ are rather weak (for $\Omega_\DE\sim 0.7$, which is closer to what we observe). However, $B$ remains an unknown constant, fundamentally, and as we pointed out earlier, it is postulated in \cite{Freivogel:2020hiz} that it may be much lower for semiclassical gravity, hence we took the liberty of setting it to a smaller value in Fig.~\ref{fig:abar} and in the left plot of Fig.~\ref{fig:OmegaDE-B} as to present more stringent constraints on $w$.

Lastly, we point out that, although throughout this section we assumed a constant DE EoS down to $a_\mathrm{eq}$, many results should be applicable if the EoS transitions to a phantom constant EoS at later times (e.g., after recombination or during the dark ages). This is due to the fact that dark energy does not significantly influence the background evolution at earlier times. Consequently, setting $\bar{a}$ after the transition period and considering the upper bound on $\Delta a$ to be approximately from the transition to today should offer a reasonable estimate for these scenarios as well.

\subsection{Varying equation of state}

In the previous subsection, we assumed a constant EoS for DE. While a constant EoS parameter is a good approximation for `perfect fluids' such as radiation ($w=1/3$), dust ($w=0$), or a cosmological constant ($w=-1$), it is harder to achieve this theoretically for phantom DE (besides perhaps with a scalar field that has an exponential potential; see, e.g., \cite{Li:2003ft}). In fact, scalar field models of DE typically have a time-dependent EoS. As we mentioned before, it is even less motivated if we consider that the NEC violation is sourced by quantum fluctuations. Therefore, a more reasonable approach is to consider DE with a time-dependent EoS that can transition to the phantom regime over time. With the goal of having a model-independent approach, various time-dependent DE parametrizations have been developed. One of the most common is the Chevallier-Polarski-Linder (CPL) parametrization \cite{Chevallier:2000qy,Linder:2002et} (see, e.g., \cite{Wetterich:2004pv,Jassal:2005qc,Barboza:2008rh} for alternatives; this is certainly not an exhaustive list), which makes the hypothesis that the variation of the EoS parameter linearly depends on the scale factor to leading order as
\begin{equation}
    w(a)=w_0+w_a(1-a)\,,\label{eq:CPL}
\end{equation}
where $w_0$ and $w_a$ are two real parameters. Here $a_0=1$ is already assumed, so $w_0=w(a_0)$ and $w_a=-\dd w/\dd a$ have the interpretation of today's value of the EoS and the negative of the EoS's derivative with respect to the scale factor, respectively.

As in the previous subsection, we make the simplifying assumption that the EoS, now \eqref{eq:CPL}, applies to DE for the whole past history of the universe. Thus, we take $p_\DE(a)=w(a)\rho_\DE(a)$, and the DE's conservation equation, $\dot\rho_\DE+3H(\rho_\DE+p_\DE)=0$, can be rewritten as $\dd\ln\rho_\DE/\dd\ln a=-3(1+w(a))$, hence after integration one finds
\begin{equation}
    \rho_\DE(a)=\frac{3H_0^2\Omega_\DE}{8\pi\GN}a^{-3(1+w_0+w_a)}e^{-3w_a(1-a)}\,.\label{eq:rhoofaCPL}
\end{equation}
The relevant quantity to integrate in the SNEC \eqref{eq:SNECcomponents0}, which we recall can be expressed as
\begin{equation}
    \int_{a_\mathrm{i}}^{a_\mathrm{f}}\frac{\dd a}{H(a)}\,f_\tau^2(\lambda(a))\frac{\rho_\DE(a)+p_\DE(a)}{a^2}\geq-\frac{B}{\GN\tau^2}\,,\label{eq:SNECCLP}
\end{equation}
is thus straightforward to compute from $\rho_\DE(a)+p_\DE(a)=\rho_\DE(a)(1+w(a))$ using \eqref{eq:CPL} and \eqref{eq:rhoofaCPL}. The last ingredient needed is $H(a)$, which also allows one to compute $\lambda(a)$ from \eqref{eq:lambdaofagen}, and it follows from Friedmann's constraint equation for a mix of dust and DE,
\begin{equation}
    \frac{H(a)^2}{H_0^2}=\frac{1-\Omega_\DE}{a^3}+\frac{\Omega_\DE}{a^{3(1+w_0+w_a)}e^{3w_a(1-a)}}\,.
\end{equation}

\begin{figure}[t]
    \centering
    \includegraphics[width=0.99\linewidth]{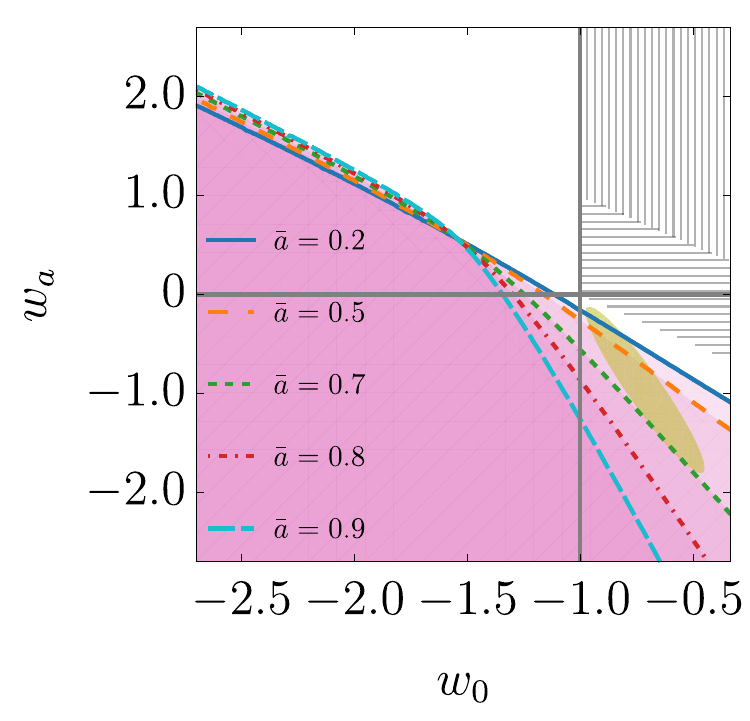}
    \caption{SNEC constraints on varying-EoS phantom (or quintom) DE. The constraint on the CPL EoS parameters $w_0,w_a$ is shown for different values of $\bar a$ (related to the center of the smearing). The other parameters are fixed to $\Delta a=a_0-a_\mathrm{eq}$, $\Omega_\DE=0.7$, and $B=1/(96\pi)$. The pink regions are where the SNEC is violated. The vertically dashed region shows where in $\{w_0,w_a\}$-space DE never gets to dominate the universe, and the horizontally dashed region represents the $\{w_0,w_a\}$-space where the classical NEC is always satisfied (and thus where the SNEC is also necessarily always satisfied). The olive ellipse gives a rough representation of current constraints from DESI \cite{DESI:2024mwx,DESI:2024hhd,DESI:2025zgx,DESI:2025kuo} when combined with CMB and supernova surveys. See text for more details.}
    \label{fig:varyingEoS_abar}
\end{figure}

The strategy to check whether the SNEC is satisfied is the same as in the previous subsection: we fix $a_\mathrm{i}=a_\mathrm{eq}$, $a_\mathrm{f}=a_0$, and use a Gaussian smearing function, whose center $\bar\lambda$ and width $\tau$ are controlled by $\bar a$ and $\Delta a$, respectively, as explained in the previous subsection. We then numerically evaluate the integrals involved in the inequality \eqref{eq:SNECCLP} at fixed parameters $\Omega_\DE$ and $B$. The results are first shown in Fig.~\ref{fig:varyingEoS_abar}, where $\Omega_\DE=0.7$, $B=1/(96\pi)$, and where furthermore we fix $\Delta a$ to its `largest allowed value', $\Delta a=a_0-a_\mathrm{eq}$, since this yields the tightest, most conservative constraint (recall the discussion in the previous subsection). The curves of different styles in Fig.~\ref{fig:varyingEoS_abar} depict the boundary of the SNEC for different choices of $\bar a$ as a function of the CPL EoS parameters $w_0$ and $w_a$; on the pink side of the boundary, the SNEC is violated, while on the white side it is satisfied. One should `ignore' the dashed regions that depict where the classical NEC is always satisfied, hence where the SNEC is trivially satisfied. In fact, in the vertically dashed region (where $w_0+w_a>0$ and $w_0>-1$), DE never really dominates the universe, and in the horizontally dashed region (where $-1<w_0+w_a<0$ and $w_0>-1$), DE eventually dominates the universe, but it is never of phantom type. (See, e.g., \cite{Chiba:2005tj,Barger:2005sb} for more refined presentations of the various subregions in the $\{w_0,w_a\}$ space.)

\begin{figure*}[t]
    \centering
    \includegraphics[width=0.45\textwidth]{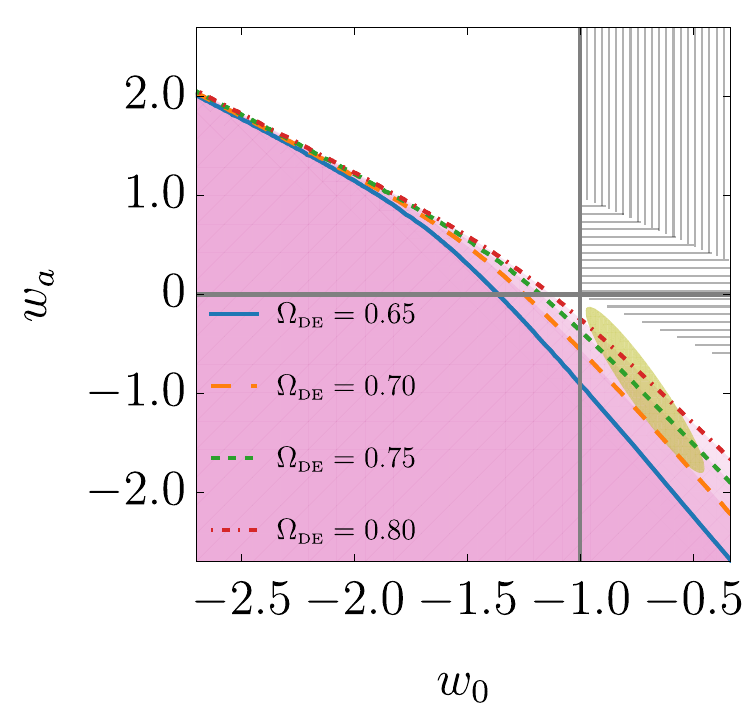}
    \hspace*{0.05\textwidth}
    \includegraphics[width=0.45\textwidth]{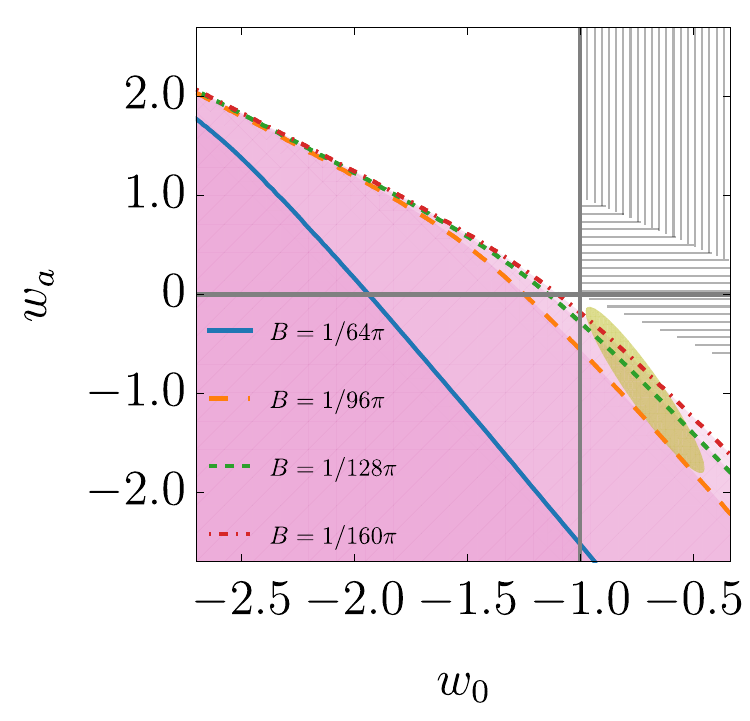}
    \caption{Same plots as Fig.~\ref{fig:varyingEoS_abar}, except now in the left plot we fix $\bar a=0.7$, $B=1/(96\pi)$, and show different values of $\Omega_\DE$; and in the right plot we fix $\bar a=0.7$, $\Omega_\DE=0.7$, and show different values of $B$.}
    \label{fig:varyingEoS_OmegaDE-B}
\end{figure*}

To further navigate the plot, note that $w$CDM (the constant EoS case analyzed in the previous subsection) is recovered along the $w_a=0$ horizontal solid gray line, and likewise, $\Lambda$CDM is recovered where this intersects the $w_0=-1$ vertical solid gray line. To give an idea of what values of $w_0$ and $w_a$ data currently favors, we added an olive ellipse, which is meant to roughly correspond to the current observational constraints at $\gtrsim 95\%$ confidence level from DESI's baryonic acoustic oscillations (BAO) when combined with type-Ia supernova (SNIa) data and cosmic microwave background (CMB) data (see \cite{DESI:2024mwx,DESI:2024hhd,DESI:2025zgx,DESI:2025kuo}).\footnote{Our olive ellipse does not correspond to any of the actual posteriors from DESI's analysis. Rather, we are showing an approximate contour that encompasses the three different combinations of BAO+CMB+SNIa data that DESI reports in \cite{DESI:2025zgx}.} While such a constraint comes with various caveats (see, e.g., \cite{Cortes:2024lgw,Linder:2024rdj,Notari:2024zmi,Giare:2025pzu,Sousa-Neto:2025gpj}), there nevertheless is growing evidence for evolving DE, moreover potentially crossing the phantom divide in the past (further see, e.g., \cite{Brout:2022vxf,DES:2024jxu,DES:2024fdw,DES:2025bxy,Chan-GyungPark:2024mlx,Huang:2024qno,Giare:2024ocw,Wang:2025ugc,Lu:2025gki}), hence we use DESI's results as a guiding reference for where observational constraints might lie.

Figure \ref{fig:varyingEoS_abar} shows how the center of the smearing function affects the SNEC constraint on $w_0$ and $w_a$. In the far top-left quadrant ($w_0<-1$, $w_a>0$), the resulting constraint depends marginally on $\bar a$. In that region, DE becomes phantom only close to today, so a larger $\bar a$ yields a tighter constraint (and correspondingly a smaller $\bar a$ yields a weaker constraint), but the difference is small since the smearing width is fixed at its largest with $\Delta a\sim 1$. The effect of changing $\bar a$ is more pronounced in the far bottom-right quadrant ($w_0>-1$, $w_a<0$), where DE is phantom in the far past before eventually satisfying the NEC today (a DE EoS that crosses the phantom divide is sometimes said to be of quintom type, as originally coined in \cite{Feng:2004ad}). There, the SNEC constraint is considerably stronger the farther back the smearing is centered (assuming the EoS is applicable that far in the past). Interestingly, this is the region where the current observational constraints from DESI lie.

Once again if the SNEC, as explained earlier, should hold for any choice of smearing, then one should take the union of all constraints, i.e., the whole pink region would thus violate the SNEC. According to this interpretation, most of the olive contour would fall in the SNEC-violating region, at least for $\Omega_\DE=0.7$, $B=1/(96\pi)$, and assuming $\Delta a$ can be as large as unity [the effect of these parameters is discussed below, and in fact, the DESI constraints are fully consistent with the SNEC if, e.g., $B=1/(32\pi)$]. Another interpretation is that one should choose the smearing so that it roughly represents our ability to observe the DE's `negative energy' over the given window. Also, given that the linear approximation is likely to fail at early times, a more reasonable fundamental model for the time variation of the DE EoS might have less `negative energy' in past. Taking DESI as an example, if we use the galaxy density distribution as a proxy for the `window' over which DE is observed, then we know that it peaks at around $z\lesssim 0.8$ \cite{DESI:2024uvr,DESI:2025zgx}, which corresponds to $\bar a\gtrsim 0.56$.\footnote{In fact, DESI \cite{DESI:2025kuo} finds that the energy density of DE peaks at around $z\approx 0.45$ (correspondingly $a\approx 0.69$) and that the matter-DE equality redshift is $z\approx 0.34$ (correspondingly $a\approx 0.75$).} For $\bar a$ in that ballpark, only a fraction of the olive contour violates the SNEC when $B=1/(96\pi)$. It certainly is interesting that the observational and theoretical constraints fall more or less in the same region. What this may imply will be further discussed in Sec.~\ref{sec:conclusions}, together with the accompanying caveats.

As in the previous subsection, let us isolate the effect of changing the DE fractional density today, $\Omega_\DE$, and the SNEC constant $B$ in Fig.~\ref{fig:varyingEoS_OmegaDE-B}. For concreteness, we fix $\bar a=0.7$. In the left plot, we see that, following intuition, the more DE one has in the universe today, the more SNEC violation one has, the stronger the constraints on $w_0$ and $w_a$. Likewise, we see in the right plot that a smaller value of $B$ always leads to a stronger SNEC constraint. Note that when we compute the SNEC for $B=1/(32\pi)$ with $\Omega_\DE=0.7$, we find very weak constraints on $w_0$ and $w_a$, i.e., the constraint lies outside the range that is shown for $w_0$ and $w_a$. For $B=1/(96\pi)$, though, the constraint gets very close to current observational bounds by DESI, and as the value of $B$ is continuously lowered, a greater portion of the DESI contours falls into the regime of SNEC violation. As observations keep improving in the future, it is interesting to note that one could envision to observationally probe what the smallest allowed value of $B$ is such that bestfit models are in agreement with the SNEC. This is another aspect of our results that will be further discussed in our conclusions.

\section{Bouncing cosmology}\label{sec:bounce}

Now that we have discussed NEC violation in the late universe with phantom DE, we move on to possible NEC violation in the very early universe. Inflationary cosmology is the paradigm of primordial cosmology, yet as it stands it cannot provide a complete description of the very early universe. Inflation resolves some problems of standard big bang cosmology (e.g., \cite{Brandenberger:1999sw}), but it does so without resolving the initial cosmological singularity (see, e.g., \cite{Borde:2001nh,Yoshida:2018ndv,Geshnizjani:2023hyd}). As such, nonsingular cosmological models can serve as preludes or alternatives to inflation in the very early universe, and bouncing cosmologies are a subclass of such singularity resolving models (see, e.g., \cite{Novello:2008ra,Battefeld:2014uga,Lilley:2015ksa,Brandenberger:2016vhg} for reviews).

\begin{figure}[t]
    \begin{center}
    \includegraphics[width=0.35\textwidth]{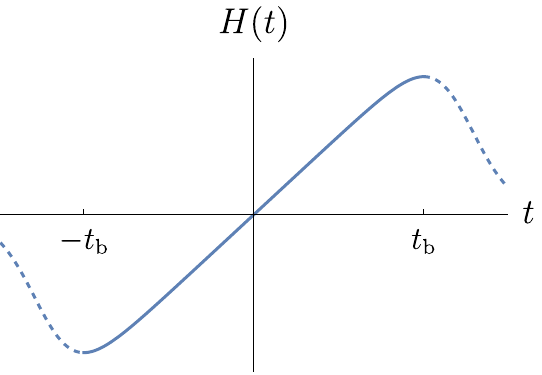}\\
    \vspace*{0.1cm}
    \includegraphics[width=0.35\textwidth]{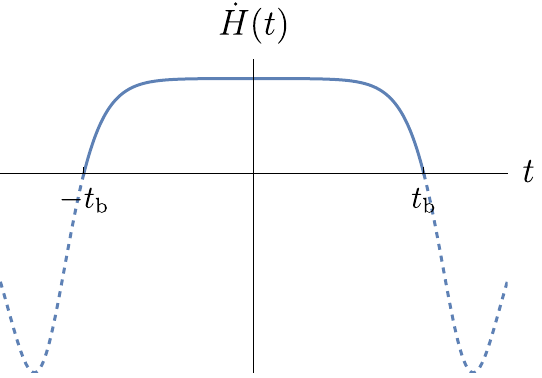}
    \end{center}
    \caption{Sketches of the Hubble parameter and its time derivative through a prototypical nonsingular bounce. The solid curve is the bounce phase during which $\dot H>0$, which is happening over the interval $t\in(-\tb,\tb)$. The dashed curve would correspond to standard NEC-satisfying cosmological evolution.}
    \label{fig:BC1}
\end{figure}

In this section, we once more focus on flat FLRW cosmology since obtaining a bouncing universe with a flat background can only be achieved with some kind of NEC-violating field(s) or states. (This is not the case with positive spatial curvature; see, e.g., \cite{Falciano:2008gt,Gao:2014hea,Bramberger:2019zez,Anabalon:2019equ,Lehners:2024qaw}.)
Indeed, we may express the flat FLRW equations with an `effective fluid' satisfying $\dot\rho_\mathrm{eff}+3H(\rho_\mathrm{eff}+p_\mathrm{eff})=0$ as
\begin{align}
    3H^2&=8\pi\GN\rho_\mathrm{eff}\,,\nonumber\\
    \dot H&=-4\pi\GN(\rho_\mathrm{eff}+p_\mathrm{eff})\,,\label{eq:FriedmannEqEff}
\end{align}
so a smooth transition from contraction ($H<0$) to expansion ($H>0$) must pass trough a `bounce point' where $H=\rho_\mathrm{eff}=0$ and where the effective NEC is violated ($\dot H>0$ and $\rho_\mathrm{eff}+p_\mathrm{eff}<0$).\footnote{Yet, the ANEC may still be satisfied --- see \cite{Giovannini:2017ndw} --- hence our main concern shall be about the SNEC.} A flat nonsingular bouncing cosmology can thus be described in terms of three phases: the prebounce contraction where $H<0$ and $\dot H\leq 0$; the bounce phase where $\dot H>0$ and where $H$ crosses zero; and the postbounce expansion where $H>0$ and $\dot H\leq 0$. See Fig.~\ref{fig:BC1} for a depiction of the Hubble parameter and its derivative, assuming a symmetric bouncing phase centered at $t=0$ (we are free to choose this origin) and where $2\tb$ represents the whole duration of the bounce phase. (More generally, it should be noted that the bounce need not be symmetrical about the bounce point, but we shall often assume so for simplicity.)

As stressed in the introduction, we put aside the challenges that one faces in trying to construct a viable NEC-violating bouncing cosmology (e.g., in terms of stability) since the focus is on exploring the implications of the SNEC if the NEC were violated (no matter how). For this reason, we simply work with a parametrization in the next subsection (as we did for DE), which is meant to describe the effective NEC-violating background evolution. However, we will look at a specific bouncing model in the subsequent subsection, the so-called Cuscuton bounce, since it is one of the most robust NEC-violating bouncing models in the literature so far.

\subsection{Parametrization}

Let us consider symmetric nonsingular bouncing models. Without loss of generality, let us center the bounce about $t=0$. We then parametrize the Hubble parameter through the effective NEC-violating bounce phase as
\begin{equation}
    H(t)=\sum_{n\in\mathbb{N}_\mathrm{odd}}\alpha_nt^n\,,\label{eq:genHparam}
\end{equation}
for some real parameters $\alpha_n$, and where $\mathbb{N}_\mathrm{odd}=\{1,3,5,7,\ldots\}$. The series expansion only contains odd powers to ensure $H(t)$ is an odd function by symmetry. Similar parametrizations have been used in the literature before (e.g., \cite{Chen:2022usd}), especially the case when $\alpha_n=0$ for all $n>1$ (i.e., keeping only a linear dependence) since it is a rather good approximation in many cases (e.g., \cite{Quintin:2014oea,Quintin:2015rta,Boruah:2018pvq,Quintin:2019orx}). One can straightforwardly take respectively a time integral and derivative of \eqref{eq:genHparam} to find the scale factor
\begin{equation}
    a(t)=\exp(\int_0^t\dd\tilde t\,H(\tilde t))=\exp(\frac{\alpha_1t^2}{2}+\frac{\alpha_3t^4}{4}+\cdots)\,,\label{eq:aBparam}
\end{equation}
and $\dot H(t)=\alpha_1+3\alpha_3t^2+\cdots$. Again without loss of generality, we parametrize the scale factor to unity at the bounce point when $t=0$. The null affine parameter $\lambda(t)$ can subsequently be computed following \eqref{eq:lambdaoftasinta}, namely $\lambda(t)=\int_0^t\dd\tilde t\,a(\tilde t)$ taking $\lambda(t=0)=0$ without loss of generality. In this case, however, we cannot write down a general, closed-form analytic expression for $\lambda(t)$. Still, we have all the ingredients needed to check the SNEC expressed as \eqref{eq:SNECtime}. Note that expressing the SNEC as such in terms of background evolution implies checking whether or not the \emph{total} SNEC is satisfied, as opposed to isolating what specific NEC-violating `quantum field' drives the bounce and checking the SNEC on that field only, as expressed in \eqref{eq:SNECcomponents01}. This is fine assuming the NEC-violating effects dominate over any NEC-satisfying matter fields present in the bounce phase and assuming they decay very rapidly outside of that phase.

Let us see why this is a good approximation. We write $\rho_\mathrm{eff}=\rho_\mathrm{s}+\rho_\mathrm{e}$ and $p_\mathrm{eff}=p_\mathrm{s}+p_\mathrm{e}$, where the labels `s' and `e' denote the `standard' (NEC-satisfying, $\rho_\mathrm{s}+p_\mathrm{s}\geq 0$) and `exotic' (NEC-violating, $\rho_\mathrm{e}+p_\mathrm{e}<0$) fields, respectively. The Friedmann equations \eqref{eq:FriedmannEqEff} then tell us that at the start and end of the bouncing phase, when $\dot H=0$, we must have $\rho_\mathrm{s}+p_\mathrm{s}=|\rho_\mathrm{e}+p_\mathrm{e}|$, indicating that the standard and exotic fields contribute equally to the total NEC at that point. Then, by the point the bounce occurs ($H=0$, $\dot H>0$), it must be that $\rho_\mathrm{e}=-\rho_\mathrm{s}$ and that $\rho_\mathrm{s}+p_\mathrm{s}<|\rho_\mathrm{e}+p_\mathrm{e}|$. While this is a minimal requirement, it is often the case that $\rho_\mathrm{s}+p_\mathrm{s}\ll|\rho_\mathrm{e}+p_\mathrm{e}|$ in order for $\dot H(t=0)$ to be sufficiently large (for the bounce phase to be sufficiently short), which motivates our assumption that the NEC-violating quantum fields dominate the bounce. Naturally, outside the bounce phase one must have the opposite limit ($\rho_\mathrm{s}+p_\mathrm{s}\gg|\rho_\mathrm{e}+p_\mathrm{e}|$) in order to approach GR with standard classical matter sufficiently quickly, hence it is reasonable to assume that the SNEC is satisfied as the NEC-violating quantum fields quickly `decay' outside the bounce phase. Altogether, this justifies the approximate equality through the bounce phase between the total SNEC \eqref{eq:SNECFLRW1} and the SNEC \eqref{eq:SNECcomponents1} applied to the exotic field(s) only.\footnote{Technically, though, this means the constraints we will find could be even stronger if the SNEC were applied to the NEC-violating content only, as we did for DE.}

The parametrization \eqref{eq:genHparam} for the Hubble parameter, or \eqref{eq:aBparam} for the scale factor, should really be thought of as a series expansion. In that sense, the Hubble parameter depends linearly on time to leading order through the bounce, so $H(t)\simeq\alpha_1t$, and $\dot H\simeq\alpha_1$ is approximately constant. From here on, let us set $\alpha_1\equiv\alpha$ to lighten the notation. Naturally, $\alpha>0$ so that $\dot H>0$. Recall that we can view the SNEC --- in a similar fashion to Heisenberg's uncertainty relation --- as an `energy-spacetime uncertainty relation', which puts an upper bound on the amount of negative-energy fluctuations of the form $\Delta E\,\Delta t\lesssim\#$, for some positive number $\#$. Through an NEC-violating bounce, we have negative-energy fluctuations $\Delta E\sim(\Delta H/\Delta t)\times\Delta t$, where $\Delta H/\Delta t\sim\dot H\sim\alpha$ and $\Delta t\sim\tb$, hence we expect the SNEC to provide a bound of the form $\Delta E\,\Delta t\sim(\Delta H/\Delta t)\times(\Delta t)^2\sim\alpha\tb^2\lesssim\#$. This makes sense if $\alpha$ and $\tb$ are the only two dimensionful parameters in the problem: the SNEC should provide a bound on the dimensionless combination $\alpha\tb^2$. Since this quantity is used repeatedly below, let us label it as $\xb\equiv\alpha\tb^2/2$ (the factor of $1/2$ simplifies the presentation below).

If we include higher-order terms in the parametrization, this introduces additional dimensionful parameters, e.g., $\alpha_3$ to next order. However, provided the parametrization is properly perturbative, the SNEC dependence on $\alpha_3$ should be small, and one should still find a bound of the form $\xb\lesssim\#$. Let us see this more explicitly. We start from the SNEC as expressed in \eqref{eq:SNECFLRW0} and consider a Lorentzian smearing function \eqref{eq:Lorentzian} for concreteness, although this can be generalized to other smearing functions. Since we center the bounce at $\lambda(t=0)=0$, we center the smearing at the same point, so $\bar\lambda=0$. The oddness of \eqref{eq:genHparam} implies the scale factor $a(t)$ and the time derivative of the Hubble parameter $\dot H(t)$ are even functions of time [and thus of $\lambda$ because $a(t)$ being an even function implies $\lambda(t)=\int\dd t\,a(t)$ is odd, hence $a(\lambda)$ is even; likewise for $\dot H(\lambda)$]. Therefore, by symmetry, we can write the SNEC as
\begin{equation}
    \int_{0}^\infty\dd\lambda\,\frac{\dot H/a^2}{\lambda^2+\tau^2/2}\leq\frac{2\sqrtb{2}\pi^2B}{\tau^3}\,.\label{eq:SNECL}
\end{equation}
If we consider \eqref{eq:genHparam} truncated to cubic order, $H(t)=\alpha t+\alpha_3t^3$, we have $a(x)=e^{x(1+\beta x)}$ and $\dot H(x)=\alpha(1+6\beta x)$, where we defined the dimensionless parameter $\beta\equiv\alpha_3/\alpha^2$, and at fixed parameter $\alpha$, we defined the rescaled dimensionless time variable $x\equiv\alpha t^2/2$. Then, from $\lambda(t)=\int_0^t\dd\tilde t\,a(\tilde t)$, we find $\lambda(x)=\mathcal{F}(x)/\sqrtb{2\alpha}$, where
\begin{equation}
    \mathcal{F}(x)\equiv\int_0^x\dd\tilde x\,\frac{e^{\tilde x(1+\beta\tilde x)}}{\sqrtb{\tilde x}}\,.
\end{equation}
From this, note $\tau=\lambda(\tb)=\mathcal{F}(x_\mathrm{b})/\sqrtb{2\alpha}$. Further note that $a(x)=1+x+(\beta+1/2)x^2+\ldots$ is a valid series expansion provided $x<1$ and $-1/x<|\beta+1/2|<1/x$. In particular, we are considering the range $0\leq x\leq x_\mathrm{b}$, so we should ask for $-1/x_\mathrm{b}<|\beta+1/2|<1/x_\mathrm{b}$ and $x_\mathrm{b} < 1$. In fact, for $x\leq x_\mathrm{b}\ll 1$ and $|\beta|\lesssim\mathcal{O}(1)$, we find $\mathcal{F}(x)\simeq 2\sqrtb{x}(1+x/3)$, with the dependence on $\beta$ appearing only at $\mathcal{O}(x^{5/2})$. Now, the SNEC \eqref{eq:SNECL} can be reexpressed as\footnote{As just argued, the parametrizaion is perturbative so long as $x\leq\xb<1$, but then we wish to integrate it for $x\in(0,\infty)$. This remains consistent provided the smearing function brings the integrand to zero sufficiently quickly for $x>\xb$.}
\begin{equation}
    \int_0^\infty\dd x\,\frac{(1+6\beta x)e^{-x(1+\beta x)}}{\sqrtb{x}(\mathcal{F}(x)^2+\mathcal{F}(x_\mathrm{b})^2/2)}\leq\frac{4\sqrtb{2}\pi^2B}{\mathcal{F}(x_\mathrm{b})^3}\,.\label{eq:SNECbbL}
\end{equation}
For completeness, the expression one finds when using a Gaussian smearing function is
\begin{align}
    \int_0^\infty\frac{\dd x}{\sqrtb{x}}&\,(1+6\beta x)\exp(-x(1+\beta x)-\frac{\mathcal{F}(x)^2}{2\mathcal{F}(x_\mathrm{b})^2})\nonumber\\
    &\leq\frac{4\sqrtb{2}\pi^{3/2}B}{\mathcal{F}(x_\mathrm{b})}\,.\label{eq:SNECbbG}
\end{align}
In either case, the result is an inequality to be solved for $x_\mathrm{b}$, whose solution will depend on $B$ (and marginally on $\beta$ due to the previous perturbative argument). Importantly, any separate dependencies on $\alpha$ or $\tb$ alone (i.e., not as the product $\alpha\tb^2$) have cancelled out. Hence, in this context, the SNEC always yields a bound of the form $x_\mathrm{b}\leq\#$, where $\#$ is a function of $B$ and $\beta$. (This somewhat similar to DE, where the SNEC implied, e.g., a lower bound on the dimensionless EoS parameter $w$, except in the case of DE there was nontrivial dependence on the smearing scale due to the more intricate functionality of the negative energy and the limits of the integration.) Based on what we will show next, we can conjecture that the resulting bound has to be of the form
\begin{equation}
    x_\mathrm{b}\big(1+\mathcal{O}(\beta x_\mathrm{b},x_\mathrm{b}^2,\beta x_\mathrm{b}^2)\big)\leq 2\pi B\,,\label{eq:SNECbgr}
\end{equation}
when assuming $x_\mathrm{b}\ll 1$ and $|\beta|\lesssim\mathcal{O}(1)$.

Further insight can be gained from analyzing the linear case $H(t)= \alpha t$, which in terms of \eqref{eq:genHparam} means that we set $\alpha_1 = \alpha$ and all other $\alpha_n=0$, $n>1$. In this case, $\dot{H}=\alpha$ is just a constant. Integrating the Hubble function returns $a(t)=\exp(\alpha t^2/2)$, and the integrand of the SNEC \eqref{eq:SNECtime} involves $\dot H/a=\alpha\exp(-\alpha t^2/2)$. To gain some analytical intuition about the SNEC in a bouncing universe, let us take the following rough smearing function,
\begin{equation}
    f_\tau^2(\lambda(t))=\begin{cases}
        1/(2\tau) & |t|\leq\tb \\
        0 & |t|>\tb\,,
    \end{cases}
\end{equation}
i.e., we take a continuous uniform distribution similar to \eqref{eq:CUD}, which is a positive constant during the bounce phase and zero otherwise. Such a smearing function satisfies the required normalization,
\begin{align}
    \int_{-\infty}^\infty\dd\lambda\,f_\tau^2(\lambda)&=\int_{-\infty}^\infty\dd t\,a(t)f_\tau^2(\lambda(t))\nonumber\\
    &=\frac{1}{2\tau}\int_{-\tb }^{\tb }\dd t\,a(t)=1\,,
\end{align}
where the last equality follows from taking
\begin{align}
    \tau&=\frac{\lambda(\tb)-\lambda(-\tb)}{2}=\frac{1}{2}\int_{-\tb}^{\tb}\dd t\,a(t)=\frac{1}{2}\int_{-\tb}^{\tb}\dd t\,e^{\alpha t^2/2}\nonumber\\
    &=\sqrtb{\frac{\pi}{2\alpha}}\mathrm{erfi}\left(\tb\sqrtb{\frac{\alpha}{2}}\right)\simeq\tb\left(1+\frac{\alpha}{6}\tb^2+\cdots\right)\,,
\end{align}
which involves the imaginary error function $\mathrm{erfi}$, whose approximation holds when the parametrization is perturbative, i.e., when $\sqrtb{\xb}=\tb\sqrtb{\alpha/2}\ll 1$. Then, the SNEC integral becomes
\begin{align}
    \int_{-\infty}^\infty\dd t&\,f_\tau^2(\lambda(t))\frac{\dot H}{a}=\frac{\sqrtb{\alpha/(2\pi)}}{\mathrm{erfi}\left(\tb\sqrtb{\alpha/2}\right)}\int_{-\tb}^{\tb}\dd t\,\alpha e^{-\alpha t^2/2}\nonumber\\
    &=\alpha\frac{\mathrm{erf}\left(\tb\sqrtb{\alpha/2}\right)}{\mathrm{erfi}\left(\tb\sqrtb{\alpha/2}\right)}
    \simeq\alpha\left(1-\frac{\alpha}{2}\tb+\cdots\right)\,.
\end{align}
Expressed in terms of $\xb\equiv\alpha\tb^2/2$, the SNEC \eqref{eq:SNECtime} thus reduces to
\begin{equation}
    \xb\left(1+\mathcal{O}(\xb^2)\right)\lesssim 2\pi B\,.\label{eq:boundBounce1}
\end{equation}
The calculation can be generalized upon reintroducing $\beta$, resulting in the similar looking perturbative bound \eqref{eq:SNECbgr}.

\begin{figure*}[t]
    \centering
    \includegraphics[width=0.45\textwidth]{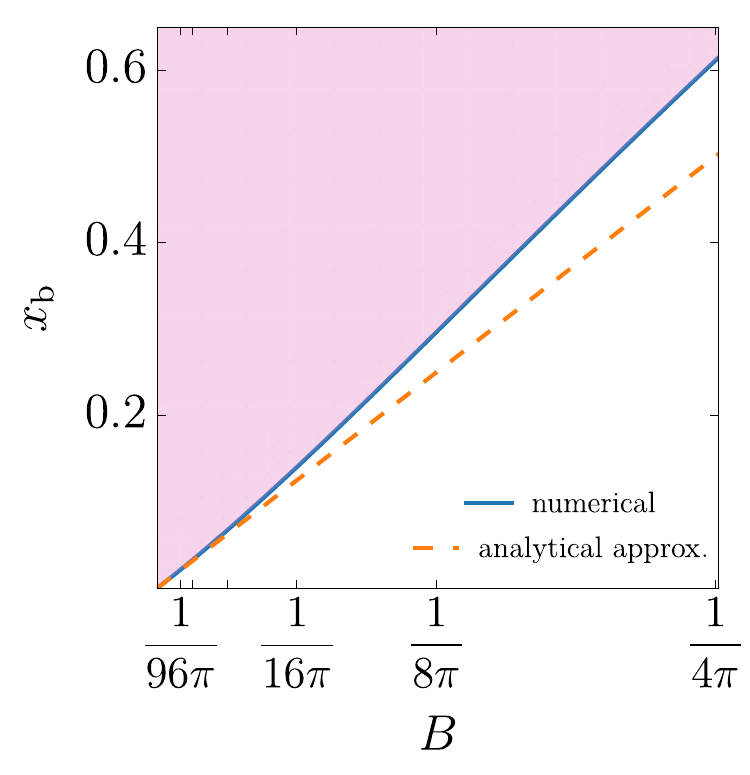}
    \hspace{0.05\textwidth}
    \includegraphics[width=0.45\textwidth]{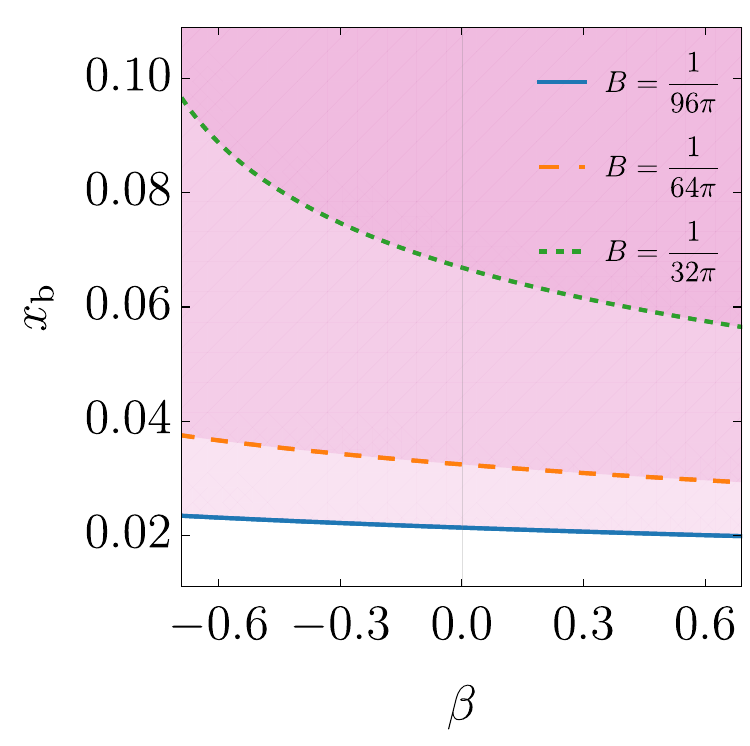}
    \caption{\textit{Left:} SNEC constraint on the dimensionless bounce-phase parameter $\xb\equiv\alpha\tb^2/2$ as a function of the SNEC parameter $B$. Ticks on the horizontal axis are depicted for $1/(2^n\pi)$, $n\in\{2,3,4,5,6\}$, and for $1/(3\cdot32\pi)$, as representative values. The pink region with blue boundary shows where the SNEC is violated, using a Gaussian smearing function and a full numerical computation. In comparison, the dashed orange line shows the analytical approximation when \eqref{eq:boundBounce1} is saturated to leading order. \textit{Right:} SNEC constraint on $\xb$ as a function of the higher-order bouncing parameter $\beta\equiv\alpha_3/\alpha^2$. Different curves depict different values of the SNEC parameter $B$.}
    \label{fig:xbvsBvsgamma}
\end{figure*}

If we use $B=1/(32\pi)$ as a representative example, \eqref{eq:boundBounce1} gives us the approximate bound $\xb\lesssim 1/16$, or equivalently, $\alpha\tb^2\lesssim 1/8$. The interpretation is once again quite straightforward: the SNEC puts an upper bound on the duration of the NEC-violating phase, $\tb$, and on the rate of NEC-violation (or the EoS), through $\alpha\approx\dot H\propto-(\rho_\mathrm{eff}+p_\mathrm{eff})$. At fixed duration, there is a lower bound on the how negative $\rho_\mathrm{eff}+p_\mathrm{eff}$ can get, or vice versa at fixed $\alpha$, there is an upper bound on the time interval of NEC violation. One could thus have, in principle, just as well a short, highly energetic bounce or a low-energy, elongated bouncing phase. To give an idea of the scales, if we take the bounce phase to be as short as a Planck time, then the SNEC implies the bound $\alpha\lesssim\mpl^2/8$, hence the Hubble scale at the start and end of the bounce phase could be very high, with $|H(\tb)|\approx\alpha\tb\lesssim\mpl/8\approx 3\times 10^{17}\,\mathrm{GeV}$. Conversely, if we say that the Hubble scale is very low, e.g., $|H(\tb)|\approx\alpha\tb\approx 100\,\mathrm{TeV}$, then the SNEC implies the bound $\tb\lesssim 3\times 10^{12}\,\mpl^{-1}$, so the bound phase could be very long (relative to a quantum gravity time scale).

The bound $\xb\lesssim 2\pi B$ comes from assuming a continuous uniform distribution for the smearing function, which allowed us to perform an analytical calculation and gain intuition. To properly apply the SNEC, the smearing function must not be nondifferentiable, though, so one should instead resort to a smooth Lorentzian or Gaussian smearing function for instance, in which case one has to solve \eqref{eq:SNECbbL} or \eqref{eq:SNECbbG} for $\xb$ numerically given values for $B$ and $\beta$. Results are shown in the left plot of Fig.~\ref{fig:xbvsBvsgamma} in the case $\beta=0$ and applying a Gaussian smearing function (very similar results are found with a Lorentzian function). Given some values of $B$, the pink region shows for what values of $\xb$ the SNEC is violated, and the boundary of that region (where the SNEC is saturated) is the solid blue line. In comparison, the line $\xb=2\pi B$ is shown in dashed orange. The approximation is good for small values of $B$ (since it corresponds to $\xb\ll 1$), but it starts to deviate from the full numerical solution, and the larger $B$ and $\xb$ are. Similar to phantom DE, the general dependence on $B$ is as expected: the smaller the value of $B$, the tighter the bound, and in fact, as $B\to 0^+$, no bounce can satisfy the SNEC as the bound on $\xb$ approaches zero (in that limit, the SNEC essentially reduces to the NEC). In contrast, a larger value of $B$ implies a weaker bound.

The case $\beta=0$ (meaning $H=\alpha t$) provides a good approximation in many bouncing instances and offers a straightforward analysis of the SNEC. However, it necessarily is an incomplete parametrization, in the sense that the cosmology can only smoothly evolve outside of the bounce phase to a NEC-satisfying regime if $\dot H$ crosses zero, as depicted earlier in Fig.~\ref{fig:BC1}. This provides motivation for exploring more general parametrizations, e.g., when at least one $\alpha_n$ is nonzero for $n>1$, but for simplicity we only explore the case $\alpha_3\neq 0$ (and we often rescale this parameter as $\beta\equiv\alpha_3/\alpha^2$). Then, we can write $H(t)=\alpha t+\alpha_3t^3$ as before, so $\dot H(t)=\alpha+3\alpha_3t^2=0$ when $t=\pm 1/\sqrtb{-3\beta\alpha}$ (assuming $\beta<0$ here). Thus, the parametrization is valid so long as $\tb\leq 1/\sqrtb{-3\beta\alpha}$ when $\beta<0$. If $\beta\geq 0$, the bounce phase `never ends' [i.e., there is no point where $\dot H(t)=0$], in which case it is the smearing function that ensures no significant negative-energy contribution is picked up for $|t|\geq\tb$. If $\beta$ is very large, though, $\dot H$ might be more important at the beginning and end of the bouncing phase, so one might have to refine the analysis further by exploring different smearing functions that better capture when NEC violation is most important.

To understand the effect of $\beta$, let us numerically solve the full expression \eqref{eq:SNECbbG} for $\xb$. We present the resulting constraint on $\xb$ in the right plot of Fig.~\ref{fig:xbvsBvsgamma} as a function of $\beta$. The pink regions are where the SNEC is violated, and the boundaries of those regions for different choices of $B$ are represented by the difference lines. We see that when $B$ is very small (so when the SNEC is very constraining), the bound on $\xb$ has little dependence on $\beta$. Nevertheless, there is a dependence [it is most obvious in the case $B=1/(32\pi)$ in the figure]: the more negative $\beta$ is, the larger $\xb$ can be, and vice versa (within the regime of validity of the approximation, which holds for the entire range shown in the right plot of Fig.~\ref{fig:xbvsBvsgamma}). This can be understood as follows: if $\beta$ is negative, it means that there is less integrated negative energy (compared to the $\beta=0$ constant case $\dot H=\alpha$), while if $\beta$ is positive, there is more integrated negative energy.

Let us end this subsection by mentioning that one could seek a `nonperturbative' parametrization of a bouncing cosmology such as $H(t)=qt/(t^2+\tb^2)$ with $q>0$. This parametrization implies $a(t)=(1+t^2/\tb^2)^{q/2}$, so near the bounce ($t\approx 0$), one recovers the linear approximation $H(t)\simeq\alpha t$ with $\alpha=q/(2\tb^2)$, hence $\xb=q/2$. Far away from the bounce, this reduces to $a(t)\sim|t|^q$, which corresponds to the expected evolution in classical GR with some matter content having an EoS $w=-1+2/(3q)$. Checking the SNEC on this parametrization, one finds the bound $q\lesssim 0.687$ [when $B=1/(32\pi)$] with practically no other separate dependence on $\tb$. In essence, one recovers that the SNEC mainly puts a constraint on the quantity $\xb$, with an upper bound here of about $0.34$. One should avoid attempting to compare this bound with, e.g., the right plot of Fig.~\ref{fig:xbvsBvsgamma}, which assumed a perturbative parametrization. Indeed, here if we expanded $H(t)$ we would find $\beta=-2/q=-\xb$, so one cannot treat $\xb$ and $\beta$ as two independent parameters as in the right plot of Fig.~\ref{fig:xbvsBvsgamma}.

Next, we consider the case of a proper bouncing cosmology model (beyond a mere parametrization) and observe the constraints that the SNEC places on the parameters of the given model.

\subsection{Cuscuton bounce}

The bouncing model we consider in this section is a product of the Cuscuton \cite{Afshordi:2006ad,Afshordi:2007yx}, which is a modified gravity theory with some interesting features. Most importantly, the Cuscuton does not introduce any additional dynamical degrees of freedom on a cosmological background. While this theory in its own right and in relation to other theories has garnered quite a lot of discussion (e.g., \cite{Afshordi:2009tt,Chagoya:2016inc,deRham:2016ged,Bhattacharyya:2016mah,Gomes:2017tzd,Boruah:2017tvg,Iyonaga:2018vnu,Lin:2017oow,Pajer:2018egx,Grall:2019qof,Mukohyama:2019unx,Aoki:2021zuy,DeFelice:2022uxv,Mylova:2023ddj}), here we restrict ourselves to the Cuscuton as providing a full example of bouncing background and explore its parameter space in light of the SNEC.

The nondynamical nature of the Cuscuton has allowed it to produce stable and robust nonsingular cosmological models (see \cite{Romano:2016jlz,Lin:2017fec,Boruah:2018pvq,Quintin:2019orx,Sakakihara:2020rdy,Kim:2020iwq,Dehghani:2025udv}), something that is typically very difficult to obtain in alternative modified gravity theories (which introduce new degrees of freedom). The readers interested in the details of this assertion are invited to explore the references above. Below we only provide a lightning review of the Cuscuton and of the corresponding bouncing model.

Before doing so, we should point out that we will apply the SNEC as expressed in \eqref{eq:SNECgeodef} to the Cuscuton bounce model. We recall that the SNEC expressed as such, despite being a geometrical condition, is motivated by a bound on NEC-violating matter in semiclassical GR. However, the Cuscuton does not straightforwardly qualify as a matter field in semiclassical GR. The situation is blurry, similar to the discussion below \eqref{eq:mod_field}. Indeed, the Cuscuton can be viewed purely as a geometrical modification to GR (still with a quantum origin; see, e.g., \cite{Afshordi:2006ad,Afshordi:2007yx,Afshordi:2009tt,Bhattacharyya:2016mah,Mylova:2023ddj}). Therefore, SNEC constraints on the Cuscuton bounce must be taken with appropriate nuances, but nevertheless, the model provides a simple, complete, and robust nonsingular background, hence it remains an obvious choice to test the applicability of the SNEC in bouncing cosmology.

The theory consists of GR, standard matter fields, and the Cuscuton modification manifested as a scalar $\varphi$. For the matter content, one can assume the presence of a massless canonical scalar field $\psi$ (i.e., $\psi$ has no potential energy, and its kinetic term is $-\partial_\nu\psi\partial^\nu\psi/2$), which would eventually decay when the universe reheats. The Cuscuton's action is given by
\begin{equation}
	S_\varphi=\int\dd^4x\,\sqrtb{-g}\left(-\mu^2\sqrtb{-\partial_\nu\varphi\partial^\nu\varphi}-V(\varphi)\right)\,,\label{eq:actionCuscuton}
\end{equation}
where $g$ denotes the determinant of the metric, $\mu$ is a new parameter, and $V(\varphi)$ is the Cuscuton's potential. Note that we set $8\pi\GN=1$ for this whole subsection. One notices that the kinetic term is noncanonical due to the presence of the square root, and it is this feature that makes the Cuscuton nondynamical when $\partial_\nu\varphi\partial^\nu\varphi<0$. (We should point out that one gets a different theory if the field's gradient is spacelike; see \cite{Afshordi:2006ad,Gomes:2017tzd,Iyonaga:2018vnu}.) The Cuscuton's action could be more general than \eqref{eq:actionCuscuton} by having an arbitrary function of $\varphi$ in front of the square root (see, e.g., \cite{Afshordi:2006ad,Mylova:2023ddj}), but the form of \eqref{eq:actionCuscuton} is ideally suited to find nonsingular cosmologies.

\begin{figure*}[t]
    \centering
    \includegraphics[width=0.45\textwidth]{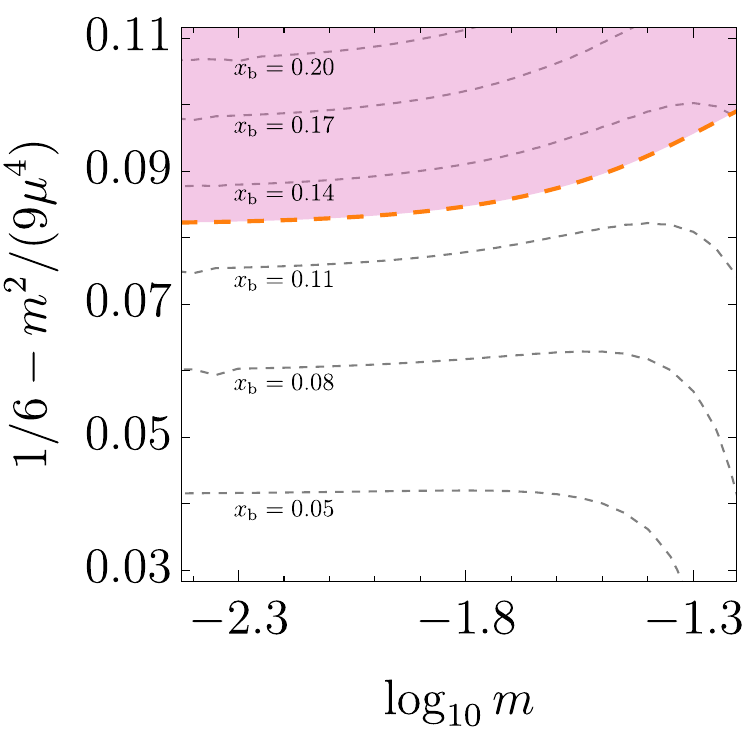}
    \hspace{0.05\textwidth}
    \includegraphics[width=0.45\textwidth]{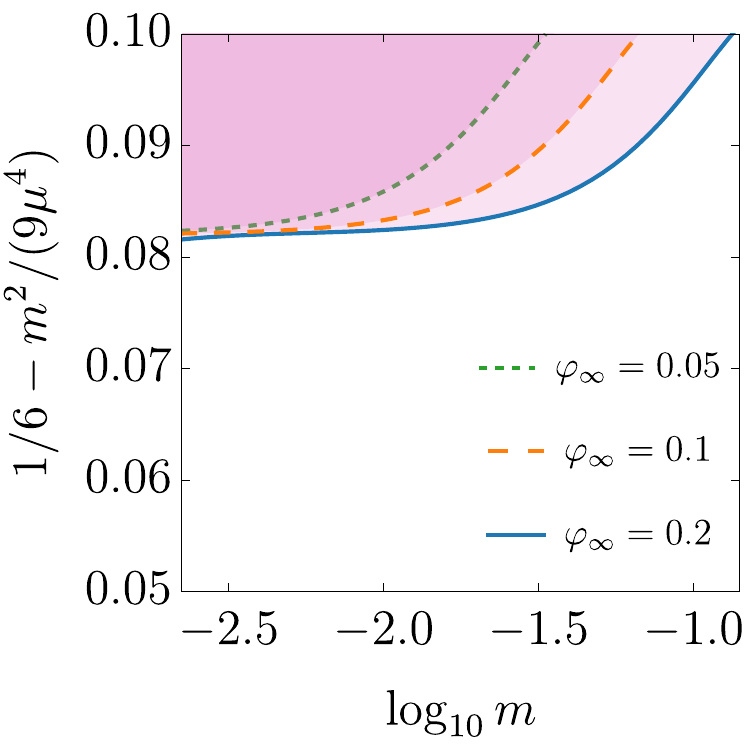}
    \caption{SNEC constraints on the Cuscuton bounce model. As before, the pink region indicates where the SNEC is violated, and the colored curves correspond to the pink regions' boundaries. The constraints on the Cuscuton bounce parameters $\{m,\mu\}$ are plotted in terms $\log_{10}m$ on the horizontal axis and $1/6-m^2/(9\mu^4)$ on the vertical axis (recall $8\pi\GN=1$ here). In the left plot, we fix $\varphi_\infty=0.1$, and in the right plot, we show the constraints for three different values of $\varphi_\infty$. Also, we assume $B=1/(32\pi)$.}
    \label{fig:Cusc}
\end{figure*}

Specializing to a flat FLRW background, the equations of motion are given by (we choose $\dot\varphi>0$ without loss of generality)
\begin{subequations}\label{eq:eomCusc}
\begin{align}
	3H^2&=\frac{\dot\psi^2}{2}+V(\varphi)\,,\\
	2\dot H&=-\dot\psi^2+\mu^2\dot\varphi\,,\label{eq:dotHeq}\\
	3\mu^2 H&=\frac{\dd V}{\dd\varphi}\,,\label{eq:constraintCuscuton}
\end{align}
\end{subequations}
in addition to the matter field's conservation equation, $\ddot\psi+3H\dot\psi=0$, which is readily solved by $\dot\psi\propto 1/a^3$. Note that \eqref{eq:constraintCuscuton} represents the Cuscuton's `evolution equation', though it truly is a constraint equation, as is evident by having no time derivatives of $\varphi$, indicating it is a nondynamical field. In fact, given a potential $V(\varphi)$ one could solve this algebraic constraint. Then, one could, in principle, reexpress all the dependence on $\varphi$ in terms of $H$ and $\dot H$ and simply obtain (complicated) modified Friedmann equations. (This shows in what sense Cuscuton gravity has more to do with a nontrivial modified gravity theory than a theory of semiclassical GR.) In practice, though, it is often simpler to solve all the equations as a differential-algebraic system.

The equations make it clear how a bounce phase can occur: when $\mu^2\dot\varphi$ is sufficiently large, it can dominate over $-\dot\psi^2$ in \eqref{eq:dotHeq} and drive a phase of effective NEC violation with $\dot H>0$. This can allow $H$ to go through zero at the point where $V(\varphi)=-\dot\psi^2/2$. One potential that can achieve this is (see \cite{Boruah:2018pvq,Dehghani:2025udv} for details)
\begin{equation}
	V(\varphi)=\frac{1}{2}m^2(\varphi^2-\varphi_\infty^2)-\frac{m^4}{4}\left(e^{2(\varphi^2-\varphi_\infty^2)/m^2}-1\right)\,,\label{eq:CuscutonPot}
\end{equation}
where $m$ and $\varphi_\infty$ are two positive constants. Provided $0<m\leq\sqrtb{3/2}\mu^2<\varphi_\infty$ (see again \cite{Boruah:2018pvq,Quintin:2019orx,Dehghani:2025udv}), a successful bounce occurs.

Our strategy is then the following: given the Cuscuton potential \eqref{eq:CuscutonPot} and some values for the parameters ($\mu$, $m$, $\varphi_\infty$), we numerically solve the system \eqref{eq:eomCusc} for $a(t)$, from which we can compute $\dot H(t)$ and $\lambda(t)$ and numerically check whether the SNEC \eqref{eq:SNECtime} is respected. As in the previous subsection when checking the SNEC numerically, we take a Gaussian smearing function centered at the bounce point and set the smearing scale to $\tau=\lambda(\tb)$. The bouncing time $\tb$ is itself found numerically from the roots of $\dot H(t)$, from which we can further find the numerical value for $\xb=\alpha\tb^2/2$ by taking $\alpha=\dot H(t=0)$.

We show our results in Fig.~\ref{fig:Cusc} assuming $B=1/(32\pi)$. To facilitate reading the constraints, we use the quantity $1/6-m^2/(9\mu^4)$ on the vertical axis. The reason is as follows: given the potential \eqref{eq:CuscutonPot}, analytical approximations from \cite{Dehghani:2025udv} suggest $H(\tb)\approx m^2\varphi_\infty/(3\mu^2)$ and $\dot H(0)\approx m^4\varphi_\infty^2/(3\mu^4-2m^2)$, hence we can estimate $\xb\approx H(\tb)^2/(2\dot H(0))\approx 1/6-m^2/(9\mu^4)$. Then, since we expect the SNEC to mainly constrain the quantity $\xb$, we plot constraints in terms of $1/6-m^2/(9\mu^4)$ as functions of $m$ (one could have equivalently chosen $\mu$ on the horizontal axis to span the $\{m,\mu\}$ space). Note, however, that $1/6-m^2/(9\mu^4)$ is not necessarily the same as the actual value of $\xb$ that we can compute numerically from the solutions. To see this, in the left plot of Fig.~\ref{fig:Cusc}, we overlay level curves of constant $\xb$ values in dashed black.

Our results confirm that the SNEC essentially provides an upper bound on the quantity $\xb$, roughly $\xb\lesssim 0.13$ when $B=1/(32\pi)$, and that this bound has marginal dependence on the various parameters. As $m$ --- or equivalently $\mu$ --- is varied by more than one order of magnitude, the bound on the ratio $m^2/\mu^4$ only changes by a small fraction. Likewise, looking at the right plot of Fig.~\ref{fig:Cusc}, we see that the bound appears to asymptote $1/6-m^2/(9\mu^4)\lesssim 0.082$ as $m$ (equivalently $\mu$) is lowered to smaller and smaller values, hence the dependence on $\varphi_\infty$ is marginal in that regime.

It is interesting to compare the bound that the SNEC puts on the Cuscuton bounce parameters with other (theoretical) constraints on the same model. For instance, it was found in \cite{Quintin:2019orx,Dehghani:2025udv} that stability and strong coupling constraints forbid the ratio $m^2/\mu^4$ from being too small or too close to $3/2$. In terms of $\xb\approx 1/6-m^2/(9\mu^4)$, this suggests $\xb$ should not be too close to zero, or too close to $1/6\approx 0.167$. Here, we find that $\xb\approx 1/6$ would certainly violate the SNEC, but having a small value of $\xb$ is consistent with the SNEC. In fact, it is interested to note that the `sweet spot' according to \cite{Quintin:2019orx,Dehghani:2025udv} where $m\approx\mu^2$ (equivalently $\xb\approx 1/18\approx 0.056$) also passes the test of the SNEC. Note that the SNEC can be even more constraining here if one takes smaller values of $B$, as we did for DE. We should end by stressing, though, that the applicability of the SNEC to the Cuscuton bounce remains a proof of principle and that any specific quantitative constraint should be taken with a grain of salt considering the fact the Cuscuton does not quite qualify as quantum matter in semiclassical GR.

\section{Discussion and conclusions}\label{sec:conclusions}

In this work, we applied the SNEC to cosmologies that would exhibit effective NEC violation. The SNEC is a proposed semilocal, semiclassical energy condition that would set a bound on how much NEC violation is physically admissible. In cosmology, there are two interesting eras that could feature matter violating the NEC: in today's DE era and in the very early universe.\footnote{In primordial cosmology, NEC violation could occur in various scenarios: bouncing cosmology, emergent cosmology, inflationary cosmology, etc. We presented an analysis on bounces, but this could certainly be extended in the future. In particular, let us mention the case of stochastic fluctuations (e.g., during inflation) that can lead to NEC violation (e.g., a scalar field `jumping' up its potential); see, e.g., \cite{Starobinsky:1986fx,Rey:1986zk,Starobinsky:1994bd}. It would thus be interesting to see how the SNEC could be applied in such a context. In particular, in the case of large fluctuations extending beyond the semiclassical regime, it might be interesting to explore whether one may define appropriate smeared spacetime energy conditions that apply to stochastic cosmological perturbation theory or full stochastic gravity (e.g., \cite{Hu:2008rga}).} In particular, phantom or quintom DE --- DE with an EoS parameter $w\ngeq-1$ at all times (i.e., it violates the NEC at some point in the history) --- has regained interest lately due to tantalizing observational evidence suggesting DE may have a time-dependent EoS, potentially crossing the phantom divide in the past (e.g., \cite{DESI:2025kuo}).

If the SNEC were shown to hold fundamentally, then our analysis in this paper shows that it would have very interesting and important consequences for phantom or quintom DE, as well as bouncing scenarios in the early universe. That is, if the SNEC were rigorously `proven' from some fundamental principles, one could apply the SNEC to constrain or rule out cosmological models.\footnote{A similar rationale is found in the swampland program. See, e.g., \cite{Agrawal:2018own,Heisenberg:2018yae,Heisenberg:2018rdu,Heisenberg:2019qxz,Heisenberg:2020ywd} for constraints on DE in that context.} Moreover, it could be used as a way to set more well-motivated priors on the `allowed' parameter space when doing Bayesian inference (e.g., what prior to choose for $w$ or $w_0,w_a$). The idea of using fundamental concepts (e.g., unitarity, causality, etc.) to constrain theoretical models and to set informed priors has been suggested in, e.g., \cite{Noller:2018eht,Noller:2018wyv,Melville:2019wyy,Noller:2020afd,deRham:2021fpu,Traykova:2021hbr,Gsponer:2021obj,Melville:2022ykg}.

Likewise, observations can provide insights into the theory by first helping us determine whether a fundamental energy condition like the SNEC is valid, and second, if it is, offering information about its properties. For example, our preliminary results indicate that $B$ values as low as $1/(96\pi)$ can be in agreement with the DESI results, and the theoretically motivated value $B=1/(32\pi)$ is certainly consistent. Thus, by examining observational data through the lens of specific models where quantum effects lead to phantom or quintom DE, we can envision applying these models to the SNEC to determine, for instance, the minimum value of $B$ data can support.

Let us highlight that when fitting models such as $w$CDM or $w_0w_a$CDM to data in observational analysis, one typically allows the DE EoS to be of phantom type in the prior (and it is not uncommon to find a bestfit of the EoS that is phantom; see, e.g., \cite{Escamilla:2023oce,DES:2024oud}), but theoretically, it is a challenge to construct viable and consistent phantom models. Here, we did not explore building a top-down model either. Rather, we took a middle ground approach asking the question of whether phantom DE could be consistent with the SNEC. The short answer is yes when taking the $w$CDM or $w_0w_a$CDM phenomenological models as proxies for NEC violation. Our work also shows when this may no longer be true. For instance, for $\Omega_\DE=0.7$ and $B=1/(96\pi)$, the SNEC is satisfied only if $w\gtrsim -1.25$ assuming the EoS is approximately constant for most of the past cosmic history (cf.~Fig.~\ref{fig:OmegaDE-B}, where the smearing function is centered at $\bar a=0.7$). More interesting are the results for the case of a time-dependent EoS and modeling NEC violation in terms of the CPL parametrization. For example, when restricted to the region where $w_0>-1$ and $w_a<0$ and again assuming $\Omega_\DE=0.7$ and $B=1/(96\pi)$, the constraint can be expressed as $w_a \gtrsim -2.49(w_0 + 1) - 0.57$ (fitting a straight line to the dashed orange curve in the bottom-right quadrant of the left plot of Fig.~\ref{fig:varyingEoS_OmegaDE-B}). This would imply that most of the $\{w_0,w_a\}$ points that fall into DESI's 2$\sigma$-contours (when combined with CMB and SNIa data) may marginally satisfy the SNEC. We should stress again that this constraint is obtained with a relatively small value of the SNEC parameter $B$. If we take the larger value $B=1/(32\pi)$ suggested by some the explicit QFT examples, the constraint is significantly reduced, while if $B\ll 1/(96\pi)$ the bound can rule out the $\{w_0,w_a\}$ values supported by DESI results.

One weakness of our analysis as already pointed out is that we only considered DE parametrizations, as opposed to specific QFT constructions leading to phantom or quintom models. While fits to data mostly focus on phenomelogical parametrizations, it is hard to pin down the microphysics purely from parametrizations --- see, e.g., \cite{Wolf:2023uno,Wolf:2025jlc} --- and one can certainly learn a great deal by constructing actual DE models that can fit the data (see, e.g., \cite{Tada:2024znt,Berghaus:2024kra,Shlivko:2024llw,Notari:2024rti,Chudaykin:2024gol,Chudaykin:2025gdn,Wolf:2024eph,Wolf:2024stt,Andriot:2024sif,Payeur:2024dnq,Shajib:2025tpd,Zhai:2025hfi,Khoury:2025txd}). Naturally, constructing real phantom or quintom models on cosmological scales, where NEC violation arise from quantum fluctuations, may be challenging. However, as a follow-up work it would be very interesting to consider time-varying DE equations of state (such as oscillatory ones) that are better motivated from such effects and constrain those models (similar to what we did for the bounce, with parametrizations versus an actual model, specifically the Cuscuton bounce; we further discuss this below). In other words, DE parametrizations are best suited to model DE in the late universe (since this is where DE's effects are most important). However, when applying the SNEC to a DE component, we assumed the parametrization was valid essentially at all times, i.e., we extrapolated it all the way to matter-radiation equality. Of course, in the region of interest the contribution of DE to the background evolution is not significant at early times, but it determines where it becomes dominant. At the same time, the smearing function in the SNEC acts as a window function that can constrain the accumulation of NEC violation at any time, hence it makes sense to apply it to the era where NEC violation is most expected. Consequently, we argued that one should center the smearing function around the times when we best measure DE (e.g., $\bar a\sim 0.7$). Yet, we also saw that the SNEC constraints were most stringent for large smearing scales (of the order of the whole universe), hence our smearing function necessarily picked up accumulation of NEC violation from the early universe. These caveats could be bypassed entirely with a better-motivated DE model since one would (in principle) trust its EoS throughout time, and hence one could robustly apply the SNEC to it and have more control on the smearing parameters. Additionally, although smearing scales might apply up to the scales of the radius of curvature \cite{Freivogel:2020hiz}, one could argue that taking this limit is quite conservative, and so, the constraints may be weaker.

In this work, we also explored possible NEC violation in the context of nonsingular bouncing universes. We found that the SNEC translates into an upper bound on the quantity $\xb\equiv\alpha\tb^2/2$, where $2\tb$ represents the duration of the NEC-violating bounce phase and where $\alpha$ amounts to the value of $\dot H$ at the bounce point (when $H$ crosses zero). At fixed $B$, we found the bound on $\xb$ has little dependence on other model parameters. We confirmed this in the case of perturbative parametrizations of the bounce phase, as well as for a specific bouncing model, namely the Cuscuton bounce. With a simple parametrization and under some approximations, the bound can be analytically derived as $\xb\lesssim 2\pi B$, and with $B=1/(32\pi)$, this yields $\xb\lesssim 0.0625$. In comparison, the bound is $\xb\lesssim 0.13$ in the case of a Cuscuton bounce. In the latter case, the bound on $\xb$ mainly translates into constraints on the model parameters $m$ and $\mu$, which are consistent with other theoretical constraints on the model coming from stability and strong coupling considerations.

There is an interesting parallel to make in this case with \cite{Kanai:2024zsw}, which demanded a certain quantum null energy condition (specifically the quantum focusing conjecture) to fundamentally hold over a certain smeared length scale. From this, they derived the cutoff scale of a semiclassical effective theory of gravity (specifically higher-dimensional quadratic gravity). We can share this interpretation, as the bound we found on $m$ and $\mu$ from the SNEC could be translated into a strong coupling scale in light of \cite{Dehghani:2025udv}. However, it is also important to recognize that the Cuscuton is a `nonlocal' modification of GR that does not introduce any new local dynamical degrees of freedom. Therefore, the extent to which the SNEC applies to the Cuscuton bounce specifically, compared to other bouncing theories derived from local perturbative quantum gravity theories, remains uncertain. We leave these interesting questions for future work.

Nonsingular bouncing cosmologies also share features with wormholes and nonsingular black holes. Restrictions on wormholes from quantum-inspired energy conditions have already been the subject of a lot of work (let us mention \cite{Kontou:2024mtd} as only one example), but for regular black holes it deserves closer scrutiny. Exploring the consequences of the SNEC on various nonsingular black hole spacetimes (e.g., \cite{Bardeen:1968,Dymnikova:1992ux,Hayward:2005gi,Zaslavskii:2010qz,Rovelli:2014cta,Fan:2016hvf,BenAchour:2017ivq,Carballo-Rubio:2018pmi,Simpson:2018tsi,Simpson:2019mud,Carballo-Rubio:2019fnb,Kelly:2020lec,Lobo:2020ffi,Maeda:2021jdc,Ashtekar:2023cod,Rovelli:2024sjl,Davies:2024ysj,Carballo-Rubio:2025fnc,Casadio:2025pun,Borissova:2025msp}) would constitute another interesting follow-up analysis.

We would like to end by mentioning that the SNEC is not the only proposal in terms of quantum-motivated energy conditions. As one additional example, we could mention the quantum null energy condition of \cite{Bousso:2015mna}, which has been proven in various contexts \cite{Bousso:2015wca,Fu:2017evt,Balakrishnan:2017bjg,Koeller:2015qmn,Malik:2019dpg,Koeller:2017njr,Akers:2017ttv,Ecker:2019ocp,Ecker:2017jdw,Moosa:2020jwt,Hollands:2025glm}, and it would certainly be interesting to further explore its consequences on the cosmologies of this work. Nevertheless, it still seems very plausible that something akin to the SNEC might hold fundamentally (like Heisenberg's uncertainty principle), so while it might not be exactly the SNEC as considered in this paper (in fact, the SNEC itself has refined formulations \cite{Fliss:2021phs,Fliss:2023rzi,Fliss:2024dxe} that could be further explored), it might still have the correct flavor. Exoticism in the universe may have its limit after all. Still, this is not incompatible with the universe having endured some NEC fracture on cosmological scales and if anything the recent observations of \cite{DESI:2024mwx,DESI:2024hhd,DESI:2024aqx,DESI:2024kob,DESI:2025zgx,DESI:2025kuo} may already be hinting that to us. Time will certainly tell whether there really was a fracture, and if so, how swiftly the universe will recover from it.

\begin{acknowledgments}
We thank Dimitrios Krommydas, Jean-Luc Lehners, and Anna Negro for stimulating discussions. This research was supported by a Discovery Grant from the Natural Science and Engineering Research Council of Canada (NSERC). This research was also supported in part by the Perimeter Institute for Theoretical Physics. Research at the Perimeter Institute is supported by the Government of Canada through the Department of Innovation, Science and Economic Development and by the Province of Ontario through the Ministry of Colleges and Universities. J.Q.~further acknowledges financial support from the University of Waterloo's Faculty of Mathematics William T.~Tutte Postdoctoral Fellowship.
\end{acknowledgments}

\bibliographystyle{JHEP2}
\bibliography{references}

\end{document}